\documentclass[a4paper, 11pt]{article}
\usepackage{tikz}
\usepackage{fullpage}
\usepackage{amsthm, amsfonts, amsmath}
\usepackage{amssymb}
\allowdisplaybreaks

\newcommand{\cX}{\mathcal{X}}
\newcommand{\cF}{\ensuremath{\mathcal{F}}}

\newcommand{\cT}{\ensuremath{K}}
\newcommand{\cL}{\mathcal{K}}

\newcommand{\R}{\ensuremath{\mathbb{R}}}
\newcommand{\N}{\mathbb{N}}

\newcommand{\conv}{*}

\newcommand{\p}{\mathbf{P}}
\newcommand{\E}{\mathbf{E}}

\newcommand{\fl}{\mathrm{Fl}}

\newcommand{\ub}{{\scriptstyle\text{\textbf{(UB)}}}}
\newcommand{\cb}{{\scriptstyle\text{{\textbf{(CB)}}}}}
\newcommand{\indep}{{\scriptstyle(\bot\!\!\!\!\bot)}} 
\newcommand{\gs}{{\scriptstyle\text{\textbf{(GS)}}}}

\newtheorem{theorem}{Theorem}
\newtheorem{lemma}{Lemma}
\newtheorem{example}{Example}
\newtheorem{remark}{Remark}
\newtheorem{assumption}{Assumption}

\begin{document}
	
	\title{Stochastic Network Calculus with Localized Application of Martingales}

	\author{Anne Bouillard \\ Huawei Technologies France \\ {\tt anne.bouillard@huawei.com} }

	\maketitle
	\abstract{Stochastic Network Calculus is a probabilistic method to compute performance bounds in networks, such as end-to-end delays. It relies on the analysis of stochastic processes using formalism of (Deterministic) Network Calculus. However, unlike the deterministic theory, the computed bounds are usually very loose compared to the simulation. This is mainly due to the intensive use of the Boole's inequality. On the other hand, analyses based on martingales can achieve tight bounds, but until now, they have not been applied to sequences of servers. In this paper, we improve the accuracy of Stochastic Network Calculus by combining this martingale analysis with a recent Stochastic Network Calculus results based on the Pay-Multiplexing-Only-Once property, well-known from the Deterministic Network calculus. We exhibit a non-trivial class of networks that can benefit from this analysis and compare our bounds with simulation. }
	\section{Introduction}
	\label{sec:introduction}
	New communication technologies aim at providing strong end-to-end delay guarantees. Some of these guarantees can be {\em deterministic}, in the sense that all packets must be transmitted in time less than some predefined value. However, these guarantees are in some cases too conservative, as the worst-case delay of a packet can be very large and rarely occurs. To this end, performances of type "the delay of a packet must be less than 10ms in 99.99\% of the cases" is desired. Examples of applications are industrial networks, virtual reality, audio and video conferencing networks. 
	
	Network Calculus is a theory that aims at providing performance bounds, such as end-to-end delays or buffer occupancy for large classes of systems:  general classes of arrival processes and of service duration, different service policies, general topologies.
	From the seminal work of R. Cruz~\cite{Cruz1995}, Deterministic Network Calculus (DNC)~\cite{Chang2000,LT2001,BBL18} has succeeded to provide accurate performance bounds in many types of networks (aircraft, industrial network,  time-sensitive networking...). 
	Stochastic Network Calculus (SNC) takes its root in the works of Yaron and Sidi~\cite{YS93} and Chang~\cite{Chang1994}, that respectively led to the tailbounded and moment-generating-function (MGF)-based SNC.

	\paragraph{Several existing SNC analyses.}
	The {\em tailbounded} SNC, analyzed in depth in Liu and Jiang's book~\cite{LJ08}, is based on the computation of the probability of elementary events, for instance, "the arrivals during a given interval of time exceeds $b$". The probability that the end-to-end delay satisfies some deadline is then obtained by the  combination of these elementary events. The analysis of large networks requires a smart combination of these events, see e.g. by Ciucu et al.~\cite{CBL06}. Some advanced results from DNC can be combined with this framework~\cite{BN15}. As such, it can handle many scheduling policies and topologies.

	The MGF-based SNC is based on the computation of the MGF of the arrival and service processes. When all these processes are independent, then their moment generating functions can easily be combined, making aggregation of flows more accurate than with the tailbounded approach. The computation of end-to-end delays however still requires the combination of elementary events. When processes are independent, the MGF-based SNC leads to better bounds than the tailbounded SNC.  In particular, when considering a system with $n$ servers in sequence, crossed by independent flows of data, the performance bounds scale in $O(n\ln n)$~\cite{CBL06} with tailbound analysis, whereas Fidler~\cite{Fid06} proved a $O(n)$ scale with MGF analysis. A precise comparison between tailbounded and MGF-based SNC has also been done by Fidler and Rizk in~\cite{FR15}.
	
	However, this method has less modeling power, as only the {\em blind multiplexing} policy ({\em i.e.,} when the  service policy is unknown, and the worst-case is considered) has been studied until now. When flows are not independent, the Hölder inequality is used to compute performance bounds. Until recently, the analysis of tandem network also induced the application of some Hölder inequality to handle the dependency of processes after crossing a common server~\cite{NS17-1}.
	A recent work~\cite{BNS22a, BNS22b}, based on the {\em Pay-Multiplexing-Only-Once}  principle \cite{BGLT2008,SZM08},  proposes to adapt a result from DNC in this framework and get rid of the Hölder inequalities for independent processes.

	One shared drawback of these two methods is that they strongly rely on the application of the union bound (also known as Boole's inequality), which makes them highly inaccurate, even for small networks. Some improvements, like {\em flow prolongation}~\cite{NS20}, can be used to reduce the size of the network to analyze, hence improve the accuracy, but does not allow getting rid of these union bounds.
	
	In order to avoid the use of the union bound, Poloczek and Ciucu in~\cite{PC14} use a martingale representation for the processes and rely on the Doob's inequality for super-martingales. The delay bounds computed that way are almost tight compared to the simulation. This method generalizes previous works by Duffield~\cite{Duf94} and Kingman~\cite{Kin64} from queuing theory to several service policies, to very general arrival processes~\cite{CP19} and to systems with replications~\cite{CPCC21}. Most works focus on random arrivals with deterministic service. In contrast,~\cite{PC15} defines a service-martingale to extend to stochastic services with applications to wireless connections and random access protocols. 
	However, this method has been applied only to networks that have no server in sequence.

	\paragraph{Contribution of the paper.}
	The contribution of this paper is to extend the martingale analysis from Poloczek and Ciucu~\cite{PC14} to sequences of servers (called {\em tandem networks}). Since it does not seem possible to define a unique martingale representing the whole tandem network, we rather focus on the application of the martingale analysis at a server only and combine it with the recent results related to the MGF-based SNC of~\cite{BNS22a}. We show on small examples the improvements that can be obtained, as well as the limits of our approach. 
	
	Note that another analysis of tandem networks can be founded in the literature~\cite{AK11}, which is attempted to use sub-martingales for the analysis of tandem networks, and recently used for the analysis of URLLC (Ultra Reliable Low Latency Communications) networks~\cite{YCL22}. Unfortunately, this approach is not sound, and we explain why as another contribution of this paper. 
	
	\paragraph{Organization of the paper.}
	The paper is organized as follows: in Section~\ref{sec:framework}, we introduce the necessary framework: MGF-based stochastic Network Calculus, the class of stochastic models used in this paper, namely the Markov-modulated processes (MPP), and a more recent formalization of MGF-based SNC based on analytical combinatorics and corresponding results on PMOO analysis from~\cite{BNS22a}. In Section~\ref{sec:results}, we present the main result of the paper. We start by generalizing Theorem~3 from~\cite{Duf94} we present a key theorem for the local application of the Doob's martingale inequality, and then use it for the analysis of tandem networks. For the sake of clarity, we explain our approach through a toy network before giving the general result. In Section~\ref{sec:AK}, we explain why the analysis of~\cite{AK11} and~\cite{YCL22} is not sound. Finally, in Section~\ref{sec:numerical}, we compare our performance bounds with the simulation and with the bounds of~\cite{BNS22a}. Proofs of the results are given in Section~\ref{sec:proofs}.
	
	\section{Stochastic Network Calculus framework}
	\label{sec:framework}
	In this section we present the necessary framework for our analysis. First, we define the Network Calculus formalism. Second, we specialize it to the MGF-based SNC and to the class of Markov-Modulated processes. More details can be found in~\cite{Chang2000,Fid06}. Then we give a combinatorial presentation of the MGF-based SNC, that allow to present results from~\cite{BNS22a} on tandem networks.

	In the whole paper, we assume time and space are discrete. 
	We deal with bivariate functions, and always assume that their definition domain is $\N^2_{\leq} = \{(t, u)\in\N^2 ~\vert~t\leq u\}$ and that they are in the set $\cF = \{f: \N^2_{\leq} \to \N_+~\vert~\forall t\geq 0,~f(t, t) = 0\}$. The main notations, defined below,  are summarized in Table~\ref{tab:notations}.
	
	\begin{table}
		\centering
		{\footnotesize \begin{tabular}{|l|l|}
				\hline
				$t, u, v$ & time variables\\
				
				\hline
				\hline
				Flows and data processes & \\
				\hline
				$i, m$ & index of flows, number of flows \\
				$a_t$ & arrivals at time slot $t$\\
				$A$, $A_i$ & bivariate arrival processes\\
				$F_A(\theta, z)$ & arrival bounding generating function\\
				$f_i, \ell_i$ & first and last servers crossed by flow $i$ \\
				$D$, $D_i$ & bivariate departures processes\\
				\hline
				\hline
				Service processes & \\
				\hline
				$j, n$ & index of a server, number of servers\\
				$s_t$ & service at timeslot $t$\\
				$S$, $S_j$ & bivariate service processes \\ 
				$F_S(\theta, z)$ & service bounding generating function\\
				
				\hline
				\hline
				Performance bounds & \\
				\hline
				$q(t)$ & backlog at time $t$\\
				$d(t)$ & delat at time $t$\\
				$F_d(\theta, z)$ & delay bounding generating function\\
				\hline
				\hline
				Generating functions & \\
				\hline
				$F(z) = \sum_{k\in\N} f_k z^k$ & Generating function associated with sequence $(f_k)_{k\in\N}$\\
				$r_F$ & dominant singularity (or radius of convergence) of $F$\\
				\hline 
				\hline 
				Markov-modulated processes &\\
				\hline
				$\cX$, $P$, $\pi$ & state space, transition matrix, and stationary distribution\\
				$\varphi_x(\theta)$ & MGF associated to state $x$\\
				$\psi(\theta)$ & exponential transition matrix, \\  $\lambda(\theta)$, $\nu(\theta)$ &  its largest eigenvalue and  associated eigenvector\\
				$(\sigma(\theta), \rho(\theta))$ & MGF-based SNC characterization of a processes\\
				$M(\theta, u, v)$ & martingale representation of a process\\
				\hline
		\end{tabular}}
		\caption{Table of notations}
		\label{tab:notations}
	\end{table}
	\subsection{Network Calculus formalism}
	\label{ssec:nc}
	\paragraph{Arrival processes.}	A bivariate process $A\in\cF$ of a flow represents the amount of data of that flow arrived in the network during any interval of time: let $a_t\in \N$ be the amount of data arriving during the $t$-th time slot. We define for all $t\leq u$, $A(t, u) = \sum_{v=t}^{u-1} a_v$, with the convention $A(t, t) = 0$. 
	In addition to the previous assumption, $A$ is additive: $A(t, u) + A(u,v) = A(t, v)$.
	
	\paragraph{$S$-servers.} Let $S\in \cF$ be a bivariate function. A server is a dynamic $S$-server if the relation between its bivariate arrival and departure processes $A\in \cF$ and $D\in\cF$ satisfies for all $t\geq 0$, $A(0, t)\geq D(0, t) \geq A\conv S(0, t),$
	where $f\conv g(t, u) = \min_{t\leq v\leq u} f(t, v) + g(v, u)$ is the (min,plus)-convolution. 
	
	This notion of dynamic $S$-server is often too weak to perform network analysis. Indeed, it can be interpreted as the stochastic version of the {\em (min,plus) service curve} from Deterministic Network Calculus~\cite{BBL18}. In most cases, (min,plus) service curves do not allow computing a service guarantee of a flow competing with another one, and stronger versions of service guarantees, like the {\em strict service curves}, are necessary. The equivalent in Stochastic Network Calculus is work-conserving $S$-servers: define $s_t$ as the amount of service offered by the server during time slot $t$. Then the service offered by this server is $S(t, u) = \sum_{v=t}^{u-1} s_v$, which defines $S$ as an additive bivariate function of $\cF$. 
	If during the time interval $(t, u]$ the server is never empty (for all $v\in(t, u]$, $A(0, v) > D(0, v)$), then for all $t\leq v\leq u$, $D(v, u) = S(v, u)$).

	In general, a server is crossed by several flows, with respective arrival and departure processes $A_i$ and $D_i$, $i\in\{1, \ldots, m\}$ in case of $m$ flows. We say that a server  is a dynamic $S$-server (resp. a work-conserving $S$-server) if it is for the {\em aggregated} arrival and departure  processes $A = \sum_{i=1}^m A_i$ and $D = \sum_{i=1}^m D_i$.  Moreover, we assume that the system is causal, and we also have for all $i\in\{1, \ldots, m\}$ and all $t\geq 0$, $A_i(0, t) \geq D_i(0, t)$.

	\paragraph{Performance bounds.}
	Consider a dynamic $S$-server and $A$ and $D$ its respective arrival and departure bivariate processes. 
	
	The backlog at time $t$ is $q(t) = A(0, t) - D(0, t)$ and the virtual delay at time $t$ is $d(t) = \inf\{ T \geq 0~\vert~A(0,t) \leq D(0, t + T)\}$.  
	
	\begin{theorem}[Performance bounds \cite{Chang2000,Fid06}] 
		\label{th:bounds} 
		Let $A$ be a bivariate process crossing an $S$-dynamic server. Then
		\begin{itemize}
			\item $q(t) \leq \sup_{0\leq s\leq t} A(s, t) - S(s, t)$;
			\item for all $T\in\N$, $d(t) \geq T \implies \exists u\leq t,~A(u, t) > S(u,t+T - 1)$. 
		\end{itemize}
	\end{theorem}
	We call the delay {\em virtual delay} because this is the delay when data exit the system in their arrival order. This is not necessarily the case when several flows cross the same system.

	\subsection{Markov modulated processes and SNC} 
	\label{ssec:mmp}
	Stochastic Network Calculus is the study of systems described above when $A$ and $S$ are described by stochastic processes, and we are interested in bounding the {\em violation probability} of some backlog or delay. More precisely, one wishes to find upper bounds for $\p(q(t) \geq b)$  and  $\p(d(t) \geq T).$
	
	In this paper, we focus on the $(\sigma(\theta), \rho(\theta))$ representations  that enables to bound these quantities, using MGF-based network calculus. In this paragraph, we first present this $(\sigma(\theta), \rho(\theta))$ characterization, then define a family of stochastic processes, the Markov-modu\-la\-ted processes that have a $(\sigma(\theta), \rho(\theta))$ representation and will be used in this paper.

\subsubsection{MGF-based SNC and $(\sigma(\theta), \rho(\theta))$-representations}
	Moment-generating function (MGF)-based SNC mainly uses two probabilistic inequalities: the union bound (or Boole's inequality, denoted $\ub$ in the equations) and Chernoff bounds\footnote{{$\p(X\geq x)\leq \inf_{\theta > 0} \frac{\E[e^{\theta X}]}{e^{\theta x}}$. We in fact here do not optimize of $\theta$ at this step and rather use $\p(X\geq x)\leq  \frac{\E[e^{\theta X}]}{e^{\theta x}}$. The optimization will be performed at the end of the derivations.}} $\cb$, that require computing MGF of a process. For example, the backlog violation probability would be for all $\theta > 0$, 
	\begin{align*}
		\p(q(t) \geq b) & \leq \p(\exists u\leq t,~A(u, t) - S(u, t) \geq b) \\ 
		& \leq \sum_{u\leq t} \p(A(u, t) - S(u, t) \geq b) & \ub\\
		& \leq \sum_{u\leq t} \E[e^{\theta (A(u, t)   - S(u, t))}] e^{-\theta b}. & \cb
	\end{align*}

	Independence between the arrival and service processes $\indep$ is a common assumption in MGF-based SNC. If we assume this, we now need to bound the MGFs of $A(u, t)$ and $S(u, t)$, and this is the goal of the $(\sigma(\theta), \rho(\theta))$ representations. 
	
	\begin{itemize}
		\item  An arrival process $A$ is $(\sigma_A(\theta), \rho_A(\theta))$-constrained  if for all $(t, u)\in\N^2_{\leq}$, $\E[e^{\theta A(t, u)}] \leq e^{\theta(\sigma_A(\theta) + \rho_A(\theta)(u-t))}.$
		\item A service process is $(\sigma_S(\theta), \rho_S(\theta))$-constrained if for all $(t, u)\in\N^2_{\leq}$, $\E[e^{-\theta S(t, u)}] \leq e^{\theta(\sigma_S(\theta) - \rho_S(\theta)(u-t))}.$ 
	\end{itemize}
	
	\begin{remark}
		\begin{itemize}
			\item $\E[e^{\theta A(t, u)}]$ is increasing in $\theta$, but might not be finite from some value, in which case we set $\sigma(\theta) = \rho(\theta) = +\infty$;
			\item  $-\theta$ is used in the  MGF of the service processes instead of $\theta$. This is to  upper bound the arrival processes, and lower bound on the service processes. 
		\end{itemize}
	\end{remark}

	We are now ready to complete the computation of the backlog violation probability, as we obtain a geometric sum $\gs$:
	\begin{align*}
		\p(q(t) \geq b) &\leq \sum_{u\leq t} \E[e^{\theta A(u, t)}]\E[e^{-\theta  S(u, t)}] e^{-\theta b}. & \indep\\
		& \leq \sum_{u\leq t}e^{\theta(\sigma_A(\theta + \rho_A(\theta)(u-t))}  e^{\theta(\sigma_S(\theta) -  \rho_S(\theta)(u-t))}e^{-\theta b} & (\sigma, \rho)\\
		& \leq \frac{e^{\theta (\sigma_A(\theta) + \sigma_S(\theta) - b)}}{1- e^{\theta(\rho_A(\theta) - \rho_S(\theta))}}. & \gs
	\end{align*}
	
	The sum is finite if and only if $\rho_S(\theta) > \rho_A(\theta)$, and the last line's bound is valid for all $t\in\N$ in that case. 
	We say that a system is {\em stable} if for all $t$, $\E[q(t)] < \infty$. A sufficient condition in this model is the existence of a positive $\theta > 0$ such that $\rho_S(\theta) > \rho_A(\theta)$.
	
	Violation probabilities for the delay can be obtained similarly or from~\cite[Lemma 2]{BNS22a}, and 
	$\p(d(t) \geq T) \leq \frac{e^{\theta (\sigma_A(\theta) + \sigma_S(\theta) + \rho_A(\theta) - \rho_S(\theta)T)}}{1- e^{\theta(\rho_A(\theta) - \rho_S(\theta)}}.$
	
	\subsubsection{Markov modulated processes}
	\label{sec:mmp}
	Many stochastic processes have $(\sigma(\theta), \rho(\theta))$ representations. In this paper, we focus on the family of the Markov-modulated processes (MMP), and present them in this paragraph in order to present the notations used in the rest of the paper (recalled in Table~\ref{tab:notations}).

	Let us denote by $y_t$ the amount of arrival or service at time slot $t$ (this corresponds to $a_t$ or $s_t$ defined above, depending if we are considering an arrival process or a service process).  
	In short, in  a MMP,  the amount of data $y_t$ at each time slot, $t$ is generated according to a distribution that depends on a state that is described by a homogeneous discrete-time Markov chain. 
	
	\newcommand{\on}{\ensuremath{\mathtt{On}}}
	\newcommand{\off}{\ensuremath{\mathtt{Off}}}
	
	\begin{example}[Markov-modulated On-Off process]
		A classical example is the Markov-modu\-la\-ted On-Off (MMOO) process: the Markov chain has two states, \on~and \off. When in the \off~state, there is no arrival, and the amount of arrival when in the \on~state is generated according to a distribution. Conditionally to being in the \on~ state, the amount of data generated is independent of the past evolution of the process. 
		An MMOO process can be depicted as in Figure~\ref{fig:mmoo}. The curvy edges between states represent the transition probabilities, and the straight edge from the \on~state represents the amount of data generated. 
		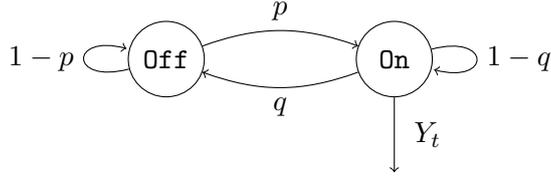
\begin{figure}[htbp]
			\centering
			\begin{tikzpicture}
				\node[circle, draw, minimum width = 1cm] (on) at (3, 0) {\on};
				\node[circle, draw, minimum width = 1cm] (off) at (0, 0) {\off};
				\draw[->] (off)  to[out=20,in=160]  node[pos=0.5, above] {$p$} (on) ;
				\draw[->] (on)  to[out=-160,in=-20]  node[pos=0.5, below] {$q$} (off) ;
				\path[->] (on) edge[loop right] node[right] {$1-q$} (on);
				\path[->] (off) edge[loop left] node[left] {$1-p$} (off);
				\draw[->] (on) -- (3, -1.5) node[pos=0.5, right] {$~Y_t$}; 
			\end{tikzpicture}
			\caption{Example of MMOO process.}
			\label{fig:mmoo}
		\end{figure}
	\end{example}

	Let us now describe an MMP more precisely, details can also be found in~\cite[Chapter 7]{Chang2000}
	Let $(X(t))_{t\in\N}$ be an ergodic discrete-time homogeneous Markov chain on a finite space $\cX$, with transition matrix $P$ and stationary distribution $\pi$. Define for all $x\in\cX$,  $(Y_x(t))_{t\in\N}$ an i.i.d sequence of random variables with moment-generating function $\varphi_x: \theta \mapsto \E[e^{\theta Y_x(0)}]$. The quantity $Y_x(t)$ represents the number of data generated  at time slot $t$ if the Markov chain is in state $x$ at time slot $t$. We assume that $(X(t))_{t\in\N}$ and $(Y_x(t))_{t\in\N}$, $x\in\cX$ are mutually independent. 
	
	\begin{example}[MMOO process (continued)]
		The Markov chain $(X(t))_{t\in\N}$ has two states: $\cX = \{\off, \on\}$. Its transition matrix and stationary distribution are given by 
		$$P = \left(\begin{array}{cc}
			1-p& p\\ q & 1-q
		\end{array}\right) \quad \text{and} \quad \pi = \left(\frac{q}{p+q}, \frac{p}{p+q}\right).$$
		There is no data generated in state \off, so $\varphi_{\off}(\theta) = 1$ for all $\theta\in\R$. If for example the data are generated according to a Poisson process of intensity $\mu$ in state \on, then $\varphi_{\on}(\theta) = e^{\mu(e^{\theta}-1)}$ for all $\theta \in\R$.
		
	\end{example}

	Then the random sequence $(y_t)_{t\in\N}$, with $y_t = Y_{X(t)}(t)$ for all $t\in\N$ is a Markov-modulated process.

	In this paper, we assume  that the process is stationary:  $X(0)$ is distributed according to the stationary distribution $\pi$ and then, for all $t\in\N$, $X(t)$ is also  distributed according to $\pi$:  $\pi P = \pi$. 
	
	As a consequence,  for all $T\in\N$, the time-reversed process $(X(T-t))_{t\leq T}$ is a homogeneous Markov chain. By Kolmogorov's extension one can extend this process to $(X(T-t))_{t\in\N}$. This process is also an ergodic homogeneous Markov chain with stationary probability $\pi$, and transition matrix $P^r = (\frac{\pi(j)}{\pi(i)}P_{j, i})_{i, j\in\cX}$. 
	
	The reversed process is often used in this type of analysis. Some authors, like Duffield in~\cite[Section 4]{Duf94}, prefer to start the analysis directly from the time-reversed process. Here we rather explicit this construction. 
	
	\noindent {\it MMOO process:} It is easy to see that a two-state ergodic Markov chain is time-reversible and that $P^r = P$. 
	
	The {\em exponential transition matrix} associated to the MMP is $\psi(\theta) = (P^r_{i, j} \varphi_j(\theta))_{i, j\in\cX}$. For all values of $\theta$ such that $\psi(\theta)$ is finite,  this matrix is primitive (as it is non-negative and irreducible). As a consequence, from the Perron-Frobenius theorem,
	\begin{itemize}
		\item $\psi(\theta)$ has an eigenvalue $\lambda(\theta)$ that is strictly positive, and strictly larger in modulus than any other eigenvalue of $\psi(\theta)$.  This eigenvalue is simple (its associated eigenspace has dimension 1);
		\item the unique right-eigenvector $\nu(\theta)$ associated to  $\lambda(\theta)$ and satisfying 
		$$\langle \nu(\theta), \pi \rangle = \sum_{x\in\cX} \pi_x \nu(\theta)_x =  1$$ 
		is strictly positive (all its coefficients are strictly positive).
	\end{itemize}

	\noindent {\it MMOO process:} For the MMOO process, we have  
	$$\psi(\theta) = \left(\begin{array}{cc}
		1-p & pe^{\mu(e^{\theta}-1)}\\ 
		q &  (1-q)e^{\mu(e^{\theta}-1)}
	\end{array}\right),$$
	the spectral analysis is left to the interested reader, that can also refer to~\cite[Example 7.2.7]{Chang2000}.

	\bigskip

	We say that an arrival process $A$  is generated by an  MMP if there exists an MMP $(a_t)_{t\in\N}$ such that  for all $(t, u)\in\N^2_{\leq}$, $A(t, u) = \sum_{i= t}^{u-1} a_i$, and similarly for a work-conserving $S$-server.
	
	In the following we will consider several arrival processes $A_i$ and servers $S_j$. The quantities defined above will be indexed by $A_i$ or $S_j$ accordingly.

	\paragraph{$(\sigma(\theta), \rho(\theta))$-characterization of a process}

	If an arrival process $A$ is generated by an MMP with exponential transition matrix $\psi_A(\theta)$, then from~\cite[Ex. 7.2.7]{Chang2000}
	\begin{itemize}
		\item $\rho_A(\theta) = \frac{1}{\theta} \ln \lambda_A(\theta)$ and 
		\item  a simple adaptation of ~\cite[Chapter 10]{Bec16} to the special case of stationary processes shows that 
		$\sigma_A(\theta) = \frac{1}{\theta} \ln \big(\frac{1}{\min_{x\in\cX} \nu_A(\theta)_x}\big).$
	\end{itemize}
	
	Similarly, if a service process $S$ is generated by an MMP with exponential transition matrix $\psi_S(\theta)$, then  $\rho_S(\theta) = - \frac{\ln \lambda_S(-\theta)}{\theta} $ and 
	$\sigma_S(\theta) = \frac{1}{\theta} \ln \big(\frac{1}{\min_{x\in\cX} \nu_S(-\theta)_x}\big).$

	\subsection{An analytic combinatorics interpretation of the $(\sigma(\theta), \rho(\theta))$- representation}
	
	In this paragraph, we use generating functions from analytic combinatorics (see~\cite{SF96} for example) also used in~\cite{BNS22a}. One advantage of using generating functions over more traditional presentations of MGF-based SNC is the compactness of the representation, but it also allows more flexibility on the characterizations of dynamic servers. Indeed, while most work use $(\sigma(\theta), \rho(\theta))$ representations to model dynamic servers, or restrict to constant-rate servers, the generating series allow representing more general characterizations, in particular when a dynamic server represents servers in tandem.
	We first give an example explaining why a service process may require a more general representation than $(\sigma(\theta), \rho(\theta))$. Then we provide a generalization that uses generating functions, and the computation of performance bounds.
	
		\subsubsection{Processes without $(\sigma(\theta), \rho(\theta))$ representation}
	
	Let us anticipate on Theorem~\ref{th:e2e-bg} stated in Paragraph~\ref{ssec:tandem}, and consider two work-conserving servers in tandem. The end-to-end dynamic server of this tandem is characterized by a product of geometric series, as we now show.  
	
	Assume $S_1$ and $S_2$ the respective service processes of the two work-conserving servers. The end-to-end service is then $\forall (u, t) \in \N_{\leq}$,
	$S_{e2e}(u, t) \geq \inf_{u\leq v\leq t} S_1(u, v) + S_2(v, t).$
	With the MGF-based analysis, we want to upper-bound $\E[e^{-\theta S_{e2e}(u, t)}]$, and we use the union bound for the upper bound of a supremum: 
	\begin{align*}
		\E\big[e^{-\theta S_{e2e}(u, t)}\big] & \leq \E\big[e^{- \theta (\inf_{u\leq v\leq t} S_1(u, v) + S_2(v, t))}\big]  
		= \E\big[\sup_{u\leq v\leq t} e^{-\theta( S_1(u, v) + S_2(v, t))}\big] \\ 
		& \leq \sum_{u\leq v\leq t} \E\big[e^{-\theta( S_1(u, v) + S_2(v, t))}\big].  \hspace{3cm}\ub  
	\end{align*}
	Now, assume that $S_i$, $i\in \{1, 2\}$, have the same $(\sigma(\theta), \rho(\theta))$ representation. Then, we can continue the computation as
	\begin{align}
		\E\big[e^{-\theta S_{e2e}(u, t)}\big] &\leq \sum_{u\leq v\leq t} e^{\theta (2\sigma(\theta) - \rho(\theta)[v-u + t-v])}  
		\hspace{3cm}(\sigma, \rho) \notag\\
		& =  \sum_{u\leq v\leq t} e^{\theta (2\sigma(\theta) - \rho(\theta)(t-u))} 
		= (t-u+1)  e^{\theta (2\sigma(\theta) - \rho(\theta)(t-u))}. \label{eq:cauchy}
	\end{align}
	We remark that here the bound given in Equation~\eqref{eq:cauchy} for the $S_{e2e}$-dynamic server is not a $(\sigma(\theta), \rho(\theta))$-representation. However, similar to the $(\sigma(\theta), \rho(\theta))$-framework, $\E[e^{-\theta S_{e2e}(u, t)}]$ depends only on the time variables by their difference, and we can express it as 
	\begin{align*}
		\E[e^{-\theta S_{e2e}(t, t+k)}] & \leq  (k+1)  e^{\theta (2\sigma(\theta) - \rho(\theta)k} = [z^k] \frac{e^{2\theta \sigma(\theta)}}{\left(1-e^{-\theta \rho_S(\theta)}z\right)^2},
	\end{align*}
{where $[z^k] G(z)$ is the $k$-th term of the generating function $G$.}
	In other words, it is possible to bound the MGF of the end-to-end service by a generating function.

	\subsubsection{Bounding generating functions}
	
	{Let us define the bounding generating functions, that allow more general representations than the $(\sigma(\theta), \rho(\theta))$ one.}
	
	\paragraph{Generating functions}
	
	Let $f = (f_k)_{k\in\N}$ be a non-negative sequence. The generating function associated to $f$ is $F(z) = \sum_{k\in\N} f_k z^k$. We use the notation $f_k = [z^k]F(z)$ to denote the $k$-th term of the sequence.  
	An important example is the geometric sequence: for all $k\in\N$, $f_k = r^k$ and $F(z) = \sum_{k\in\N} (rz)^k = (1-rz)^{-1}$.  
	
	Analytic combinatorics is a branch of combinatorics. Usually, for a collection of combinatorial objects, $f_k$ corresponds to the number of objects of size $k$. The idea is to represent this collection by the function $F(z) = \sum_{k\in\N} f_k z^k$, the generating function. The main advantage of this representation is the close relation between the asymptotic behavior of the sequence and the singularities of its generating function.
	The radius of convergence (or dominant singularity) of $F$ is defined by $r_F = \sup\{z\geq 0~|~\sum_{k\in\N} f_kz^k < \infty\}$. One important result is that if $F$ has a dominant singularity of multiplicity 1, then $f_k \sim_{k\to\infty} c r_F^{-k}$ for some constant $c$~\cite[Ch. 5]{SF96}. This can be checked for the example of the geometric series $F(z) = (1-rz)^{-1}$, whose  unique (hence dominant) singularity is $r^{-1}$.

	\paragraph{Bounding generating functions for processes}
	
	An arrival process has the arrival bounding generating function (arrival bgf) $F_A(\theta, z)$ if for all $k\in\N$, for all $t\in\N$, $\E[e^{\theta A(t, t+k)}] \leq [z^k]F_A(\theta, z)$. When $A$ has a $(\sigma_A(\theta), \rho_A(\theta))$ representation, then 
	$F_A(\theta, z) = \frac{e^{\theta \sigma_A(\theta)}}{1-e^{\theta \rho_A(\theta)}z}$
	is an arrival bgf of $A$.  Its radius of convergence is $r_A(\theta) = e^{-\theta \rho_A(\theta)}$. 
	
	Similarly, a service process has the service bounding generating function (service bgf) $F_S(\theta, z)$ if for all $k\in\N$, for all $t\in\N$,  $\E[e^{-\theta S(t, t+k)}] \leq [z^k]F_S(\theta, z)$. If $S$ is $(\sigma_S(\theta), \rho_S(\theta))$-characterized, then , $F_S(\theta, z) = \frac{e^{\theta \sigma_S(\theta)}}{1-e^{-\theta \rho_S(\theta)}z}$. Its radius of convergence is $r_S(\theta) = e^{\theta \rho_S(\theta)}$.

	\paragraph{Cauchy product of geometric series}
	In Equation~\eqref{eq:cauchy}, we recogize the Cauchy product of two sequences (the MGFs of the service processes): if $(f_k)_{k\in\N}$ and $(g_k)_{k\in\N}$ are two series, then their Cauchy product is the sequence $(h_k)_{k\in\N}$ with $h_k = \sum_{k'=0}^k f_{k'} g_{k-k'}$ for all $k\in\N$. An important result from analytical combinatorics is that the generating series of the Cauchy product of sequences is the product of the generating functions of the sequences: $H(z) =  F(z)G(z)$.
	
	Applying this to the product of $n$ geometric series translates into the following equalities: 
	for all $\alpha_1, \ldots, \alpha_n \in (0, 1)$,   
	\begin{equation}
		\label{eq:chauchy_n}
		\sum_{\substack{k_1, \ldots, k_n \geq 0 \\ k_1 + \cdots + k_n = k}} \prod_{j=1}^n \alpha_j^{k_j} = [z^k] \prod_{j=1}^n \frac{1}{1-z\alpha_j}
		\quad \text{and} \quad  
		\sum_{k_1, \ldots, k_n \geq 0} \prod_{j=1}^n \alpha_j^{k_j} = \prod_{j=1}^n \frac{1}{1-\alpha_j}.
	\end{equation}

	\subsubsection{Bounding performance using generating functions}
	Defining a delay bounding generating function (delay bgf) of a process as a generating function $F_d(z)$ such that for all $T\geq 0$, $\p(d(t) \geq T) \leq [z^T]F_d(z)$ also allows more compact representations. Since the backlog violation probability $\p(q(t) \geq b)$ has a simpler expression, we do not define one for the backlog.

	With this formalization, we can state the theorem for performance computation.

	\begin{theorem}[{\cite[Corollary 3 and Lemma 2]{BNS22a}}]
		Consider a dynamic $S$-server offering a service bounded by the service bgf $F_S(\theta, z)$ and crossed by a flow with bivariate process $A$ that is $(\sigma_A, \rho_A)$-constrained. Assume independence between the arrival and service processes. For all $\theta$ such that $r_A(\theta)r_S(\theta) > 1$, 
		\begin{enumerate}
			\item Backlog violation probability bound: $\p(q(t) \geq b)  \leq e^{-\theta b} e^{\theta\sigma_A(\theta)} F_S(\theta, e^{\theta\rho_A(\theta)}).$
			\item Delay bgf: $F_d(\theta, z) = e^{\theta \sigma_A(\theta)} \frac{e^{\theta\rho_A(\theta)}F_S(\theta, e^{\theta\rho_A(\theta)}) - z F_S(\theta, z)}{1-ze^{-\theta\rho_A(\theta)}}.$
		\end{enumerate}
	\end{theorem}

	Remind that $r_A(\theta) = e^{-\theta \rho_A(\theta)}$ and $r_S(\theta)=e^{\theta \rho_S(\theta)}$. It is shown in~\cite[Lemma 1]{Duf94} that in the stationary regime\footnote{The function $\lambda$ defined in this reference ($f$ here) is then the function $\ln \E[e^{\theta(a_1 - s_1)}]$.}, $f: \theta \mapsto \ln \E[e^{\theta(a_1 - s_1)}] = -\ln(r_A(\theta)r_S(\theta))$ is convex on its definition set, and $f(0) = 0$. Consequently~\cite[Lemma 2]{Duf94}, there exists $\theta > 0$ satisfying $r_A(\theta)r_S(\theta) > 1$ ({\em i.e.}, $f$ is decreasing on some interval $[0, x]$) if and only if $f'(0) <0$, that is $\E[a_1] - \E[s_1] < 0$: there are in average fewer arrivals than the service offered, and the system is said {\em stable}.

	We set $\theta^* = \sup\{\theta\geq 0~|~r_A(\theta)r_S(\theta) > 1\}$. This value can be $+\infty$ in case the arrivals are almost surely less than the services at each time step. 
	
	In the particular case of a $(\sigma_S(\theta), \rho_S(\theta))$-characterized dynamic server, 
	$$F_S(\theta, z) =\frac{e^{\theta\sigma_S(\theta)}}{1-e^{-\theta\rho_S(\theta)}z},$$ and the  condition $r_A(\theta)r_S(\theta) > 1$ also reads $\rho_A(\theta) - \rho_S(\theta)<0$ or $\rho_S(\theta)>\rho_A(\theta)$. The delay bgf is
	$$F_d(\theta, z) = \frac{e^{\theta(\sigma_A(\theta)+\sigma_S(\theta) + \rho_A(\theta))}}{1-e^{-\theta(\rho_S(\theta) - \rho_A(\theta))}} \cdot \frac{1}{1-ze^{-\theta\rho_S(\theta)}},$$ and the delay and backlog violation probabilities are respectively 
	\begin{align}
		\p(d(t) \geq T)& \leq \frac{e^{\theta(\sigma_A(\theta)+\sigma_S(\theta) +\rho_A(\theta)- \rho_S(\theta)T)}}{1-e^{-\theta(\rho_S(\theta) - \rho_A(\theta))}}  \quad \text{ and } \label{eq:vp-single-server-delay}\\  \p(q(t) \geq b)& \leq \frac{e^{\theta(\sigma_A(\theta)+\sigma_S(\theta) - b)}}{1-e^{-\theta(\rho_S(\theta) - \rho_A(\theta))}},\label{eq:vp-single-server-backlog}
	\end{align} for all $\theta < \theta^*.$ 
	
	In this example, we see that the violation probabilities depend on the choice of $\theta$. For a fixed value of $\theta$,  the violation probability of the backlog decreases exponentially fast with rate $\theta$. The higher the value of $\theta$, the higher the decay rate. This has to be mitigated with the denominator that decreases to 0 when $\theta$ approached $\theta^*$. Nevertheless, when optimizing the value of $\theta$ to minimize the violation probability, one can observe that $\theta$ grows to $\theta^*$ when the backlog objective grows to infinity. 
	
	\subsection{Tandem network model}
	\label{ssec:tandem}
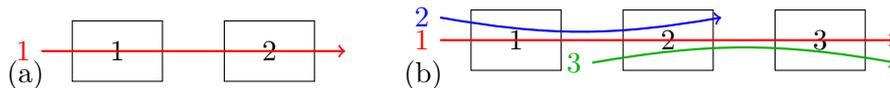
\begin{figure}[htbp]
		\centering
		(a)\hspace{-0.5cm}\begin{tikzpicture}[server/.style={shape=rectangle,draw,minimum height=.8cm,inner xsep=3ex}]
			\node[server,name=S1] at (0,0) {$1$};
			\node[server,name=S2] at (2,0) {$2$};
			\draw[->, thick, red] (-1,0) node[left] {$1$} -- (3,0);
		\end{tikzpicture}
		\hspace{0.5cm}
		(b)\hspace{-0.5cm}\begin{tikzpicture}[server/.style={shape=rectangle,draw,minimum height=.8cm,inner xsep=3ex}]
			\node[server,name=S1] at (0,0) {$1$};
			\node[server,name=S2] at (2,0) {$2$};
			\node[server,name=S3] at (4,0) {$3$};
			\draw[->, thick, red] (-1,0) node[left] {$1$} -- (5,0);
			\draw[->, thick, blue] (-1,0.3) node[left] {$2$} to[out=-10,in=-170] (2.7,.3);
			\draw[->, thick, green!70!black]  (1,-0.3) node[left] {$3$} to[out=10,in=170] (5,-.3);
		\end{tikzpicture}
		\caption{Examples of network: (a) two-server tandem; (b) interleaved tandem.}
		\label{fig:tandem}
	\end{figure}

	In this paragraph, we describe our model of tandem network. Examples are given in Figure~\ref{fig:tandem}. 
	
	\begin{assumption}[Topology]
		Let us make the three following assumptions concerning the topology of the network, and the notations will be used throughout the paper.
		\begin{enumerate}
			\item[$(H_1)$] The network is composed of $n$ servers, numbered from 1 to $n$.
			Each server $j$ is a  $S_j$-work-conserving server.
			We assume FIFO per-flow and arbitrary multiplexing: there is no assumption about the service order of data, except that data from the same flow are served in their arrival order.
			\item[$(H_2)$] There are $m$ flows circulating in the network, numbered from 1 to $m$. For all flow $i$, there exist $f_i \leq \ell_i\in\N_n$, such that flow $i$ crosses servers $f_i, f_{i}+1,\ldots, \ell_i$. We denote by $A_i$ the bivariate arrival process of flow $i$.
			We write $i\in \fl(j)$ if flow $i$ crosses server $j$.
			\item[$(H_3)$] Flow 1, for which we are going to compute performance bounds and is then also called the flow of interest, crosses the whole tandem network ($f_1 = 1$ and $\ell_1 = n$).
		\end{enumerate}
	\end{assumption}

	One technique, known as {\em Pay-Multiplexing-Only-Once} (PMOO), to compute the performance bounds of flow 1 is to first compute an end-to-end dynamic server, that represents a lower bound on the service offered to the flow and depends on the servers crossed and on the interfering flows. The formula is given by the next theorem. 
	
	\begin{theorem}[End-to-end dynamic server~{\cite[Theorem 6]{BNS21}}]
		\label{th:e2e-sc}
		With the notations and assumptions $(H_1)$--$(H_3)$, the end-to-end service offered to flow 1 is a dynamic $S_{e2e}$-server with $\forall 0\leq   t_{1} \leq t_{n+1}$,  
		$$S_{e2e}(t_{1}, t_{n+1}) =  \Big[\inf_{\substack{~t_j \leq t_{j+1}, \\1\leq j\leq n}} ~\sum_{j=1}^n [S_j(t_j, t_{j+1})] - \sum_{i=2}^m A_i(t_{f_i}, t_{\ell_i + 1})\Big]_+,$$
		where $[\cdot]_+ = \max(\cdot, 0)$.
	\end{theorem}
	
	This theorem is an adaptation to bivariate processes of the {\em Pay-multiplexing-only-once} principle from Deterministic network Calculus~\cite{BGLT2008, SZM08}. Intuitively, the end-to-end service is obtained by removing the cross traffic where it interferes with the flow of interest (flow 1). 
	
	\begin{example}
		Dynamic end-to-end servers for the networks (a)
		and (b) of Figure~\ref{fig:tandem} are: 
		\begin{itemize}
			\item 	$S^{(a)}_{e2e} (t_1, t_3) = [\inf_{t_1 \leq t_2 \leq t_3} S_1(t_1, t_2) + S_2(t_2, t_3)]_+$ and
			\item $S^{(b)}_{e2e} (t_1, t_4) = [\inf _{t_1 \leq t_2 \leq t_3 \leq t_4} S_1(t_1, t_2) + S_2(t_2, t_3) + S_3(t_3, t_4) - A_2(t_1, t_3) - A_3(t_2, t_4)]_+$. 
		\end{itemize}
		In the latter example, the end-to-end service considers the service of server $j$ on the time interval $[t_j, t_{j+1}]$. The arrival process of flow 2 is taken into account on the time interval $[t_1, t_3]$, which corresponds to the periods concerning servers 1 and 2, which are precisely the servers on the path of flow 2. 
	\end{example}
	\begin{assumption}[Processes]
		In the whole paper, we make the following assumptions on the processes:
		\begin{itemize}
			\item[$(H_4)$] arrival processes $(A_i)_{i=1}^m$ and service processes $(S_j)_{j=1}^n$ are stationary MMPs, with notations explained in Paragraph~\ref{ssec:mmp};
			\item[$(H_5)$] arrival processes $(A_i)_{i=1}^m$ and service processes $(S_j)_{j=1}^n$  are mutually independent.
		\end{itemize}
	\end{assumption}
	
	We now can give a bound of the end-to-end service using a service bgf.
	\begin{theorem}[End-to-end service bgf~\cite{BNS22a}]
		\label{th:e2e-bg}
		With notations and assumptions $(H_1)$--$(H_5)$, a service bgf for the end-to-end server of flow 1 is 
		$$F_{S_{e2e}}(\theta, z) = e^{\theta(\sum_{i=2}^m \sigma_{A_i}(\theta) + \sum_{j=1}^n \sigma_{S_j}(\theta))} \prod_{j=1}^{n}\frac{1}{1 - e^{-\theta (\rho_{S_j}(\theta) - \sum_{i\in \fl(j)\setminus\{1\}}\rho_{A_i}(\theta))}z}.$$
	\end{theorem}
	
	Theorem~\ref{th:e2e-bg} is obtained from Theorem~\ref{th:e2e-sc} by bounding the MGF of the end-to-end service $\E[e^{-\theta S_{e2e}(t, t+k)}]$ for each interval length $k$. MGF-based SNC uses the union bound to replace the expectation of a maximum (the infimum transforms into a supremum because of the "$-\theta$" MGF of a service process) as a sum of the expectations. The sum of products of expectations then translate into Cauchy products of series, and the product of the generating functions (see Equation~\eqref{eq:chauchy_n}). 
	
	\begin{example}
		Service bgf for the end-to-end servers of networks (a) and  (b) of Figure~\ref{fig:tandem} are (dependencies in $\theta$ are omitted in the second example for the sake of concision): 
		\begin{itemize}
			\item 	$F_{S^{(a)}_{e2e}}(\theta, z)  = F_{S_1}(\theta, z) \cdot F_{S_2}(\theta, z) = \frac{e^{\theta (\sigma_{S_1}(\theta) + \sigma_{S_2}(\theta))}}{\big(1-e^{-\theta\rho_{S_1}(\theta)}z\big) \big(1-e^{-\theta\rho_{S_2}(\theta)}z\big)} $ is the product of the two service bounding generating functions of the work-conserving servers and
			\item $F_{S^2_{e2e}}(\theta, z)  = \frac{e^{\theta (\sigma_{S_1} + \sigma_{S_2} + \sigma_{S_3} + \sigma_{A_2} + \sigma_{A_3})}} { \big(1-e^{-\theta(\rho_{S_1} - \rho_{A_2})}z\big)\big(1-e^{-\theta(\rho_{S_2} - \rho_{A_2} - \rho_{A_3})}z\big) \big(1-e^{-\theta(\rho_{S_3} - \rho_{A_3})}z\big)}$ can be inter\-pre\-ted as the product of the three {\em residual servers}, when the arrival processes of the cross traffic is removed from the service.
		\end{itemize}
	\end{example}

	\paragraph{Stability and bottlenecks}
	Consider a tandem network  described as above. This network is stable if the backlog of flow 1 can be bounded in expectation. In our setting, this corresponds to the existence of $\theta>0$ such that  $\forall b\in \N$, $\p(q(t) > b) \leq C(\theta)e^{-\theta b}$ for some constant $C(\theta)\in\R_+$. 
	
	When Theorem~\ref{th:e2e-bg} is combined with the arrival curve for flow 1, the backlog violation probability bound is 
	$$\p(q(t) \geq b) \leq \frac{e^{-\theta b} e^{\theta(\sum_{i=1}^m \sigma_{A_i}(\theta) + \sum_{j=1}^n \sigma_{S_j}(\theta))}}{\prod_{j=1}^n\big(1-e^{-\theta(\rho_{S_j}(\theta) - \sum_{i\in\fl(j)} \rho_{A_i}(\theta))}\big)},$$
	and is defined for all $\theta$ such that $r_A(\theta)r_{S_{e2e}}(\theta) > 1$.
	Here, $r_A(\theta) = e^{-\theta \rho_A(\theta)}$ and the dominant singularity of $F_{e2e}(\theta, z)$ is 
	$$r_{S_{e2e}}(\theta) = \min_{j\in\{1, \ldots, n\}} e^{\theta(\rho_{S_j}(\theta) - \sum_{i\in \fl(j)\setminus\{1\}}\rho_{A_i}(\theta))}.$$ 
	
	In other words, for all server $j$, one must have $\rho_{S_j}(\theta) > \sum_{i\in\fl(j)} \rho_{A_i(\theta)}$.
	Let us denote for each server $j$, $$\theta^*_j = \sup\big\{\theta \geq 0~|~\forall j\in\{1, \ldots, n\},~\rho_{S_j}(\theta)  > \sum_{i\in \fl(j)} \rho_{A_i}(\theta)\big\}.$$
	The formula for the backlog violation probability is then valid for all $\theta < \theta^* = \min_{j=1}^n \theta^*_j$. We call the {\em bottleneck(s)} the server(s) at which this minimum is reached.

	\section{Tandem analysis with localized use of martingales}
	\label{sec:results}
	In this section, we describe the main contribution of the paper. We extend to the case of multiple servers the use of martingale in the network calculus framework. As stated in the introduction, the use of the martingale is  localized at one server. Nevertheless, performance bounds are improved. In this analysis, we rather follow the approach of Duffield's paper~\cite{Duf94}. First, in Paragraph~\ref{sec:mart}, we generalize Theorem~3 of~\cite{Duf94} to Theorem~\ref{th:martingale} that will be key for the analysis. Then in Paragraph~\ref{sec:analysis}, we detail the analysis for the two-server tandem of Figure~\ref{fig:tandem}(a), that is representative enough of the approach.  Finally, we will state the general result, Theorem~\ref{th:main} in Paragraph~\ref{ssec:general-result}. 
	
	\subsection{Martingales for arrival and service processes}
	\label{sec:mart}
	In this paragraph, we start from Duffield's analysis of~\cite{Duf94} to build the martingale used for our main result and give the key theorem for the analysis of tandem networks in Paragraph~\ref{ssec:general-result}.

	Consider an arrival process $A$ generated by a MMP. Using the notation of Paragraph~\ref{ssec:mmp},  we can define for all $(u, v)\in\N_{\leq}^2$,  
	$$M_A(\theta, u, v) = e^{\theta A(u, v) - \theta\rho_A(\theta)(v-u)} \nu_A(\theta)_{X_A(u)}.$$
	Let $\cF_A(u, v)$ be the $\sigma$-algebra generated by $(X_A(u), \ldots, X_A(v), a_u, \ldots, a_{v-1})$. 
	\begin{lemma}
		\label{lem:duffield}
		For all $\theta \in \R$ such that $\psi_A(\theta)$ is defined, for all $(u, v)\in \N_{\leq}^2$, 
		$\{M_A(\theta, u-\tau, v)\}_{\tau \in \N}$ is a martingale with respect to the filtration $(\cF_A(u-\tau, v))_{\tau \in\N}$. 
	\end{lemma} 

This result is almost a rewriting of  Lemma~1 of~\cite{Duf94} to the specific case of MMP, and the complete proof can be found in Paragraph~\ref{sec:proofl1}. The main difference is that we use bivariate functions for the arrivals, and then make explicit that the martingale is used in reversed-time. If $A$ has  i.i.d increments, then $M_A(\theta, u, v) = e^{\theta A(u, v) - \theta\rho_A(\theta)(v-u)}$.

	Similarly, if $S$ is the service process of a work-conserving server generated by a MMP, we can define for all $(u, v)\in\N_{\leq}$, $$M_S(\theta, u, v) = e^{-\theta S(u, v) + \theta\rho_S(\theta)(v-u)} \nu_S(-\theta)_{X_S(u)}.$$
	If $\cF_S(u, v)$ is the $\sigma$-algebra generated by $(X_S(u), \ldots, X_S(v), s_u, \ldots, s_{v-1})$, then for all $\theta \in \R$ such that $\psi_S(-\theta)$ is defined, for all $(u, v)\in \N_{\leq}^2$, 
	$\{M_S(\theta, u-\tau, v)\}_{\tau \in \N}$ is a martingale with respect to the filtration $(\cF_S(u-\tau, v))_{\tau \in\N}$. 
	
	\begin{theorem}
		\label{th:martingale}
		Consider mutually independent arrival processes $(A_i)_{i=1}^m$ and service process $S$ {satisfying Assumptions $(H_4)-(H_5)$}, and $(u_i, v_i)_{i=0}^m \in (\N^2_{\leq})^{m+1}$ and $Y$ a random variable independent of $(A_i)_{i=1}^m$ and $S$. Define for all $\tau \in \N$,
		$$W_{\tau} = \sum_{i=1}^m A_i(u_i-\tau, v_i) - S(u_0 - \tau, v_0).$$
		For all  $\theta\in\R_+$ satisfying $\sum_{i=1}^m\rho_{A_i}(\theta) - \rho_{S}(\theta) \leq 0$, there exists a constant $\xi_{(A_i), S}(\theta)$ independent of $(u_i, v_i)_{i=0}^m$ such that 
		$$\p(\sup_{\tau\geq 0} W_\tau \geq Y) \leq \xi_{(A_i), S}(\theta) \E[e^{-\theta Y}] e^{\theta  (\sum_{i=1}^m \rho_{A_i}(\theta)(v_i-u_i) - \rho_{S}(\theta)(v_0 - u_0)))}.$$
	\end{theorem}

Compared to Theorem~3 of~\cite{Duf94}, we clearly separate the arrival processes and the service processes. This enables us to consider them from different end point: while in~\cite{Duf94}, the starting time is 0 and backward processes are implicitly used, the equivalent would be to set $u_i = v_i = 0$ for all $i\in\{0, \ldots, m\}$. So here, we introduce more flexibility in the definition of the process $(W_\tau)_{\tau\in\N}$. Nevertheless, the proof of Theorem~\ref{th:martingale} follows the lines of Theorem~3 of~\cite{Duf94}. It is given in Paragraph~\ref{app:martingale}.

{The constant $\xi_{(A_i), S}(\theta)$ depends on the MMP of the arrival and service processes. It can be expressed as $\xi_{(A_i), S}(\theta) = (\inf\{\nu(\theta)_x~|~ x\in \mathcal{P})^{-1}$, where $\nu(\theta)$ is the tensor product of the $\nu_X(\theta)$'s, $X\in\{A_1, \ldots, A_m, S\}$ and $\mathcal{P}$ is the set of states that have a positive probability to receive more arrivals that the amount of service offered. When all processes have i.i.d. increments, then $\xi_{(A_i), S}(\theta)=1$. More generally, with the choice of $\sigma_X(\theta)$ in Paragraph~\ref{ssec:mmp}, we have $\xi_{(A_i), S}(\theta) \leq e^{\theta (\sigma_S(\theta) + \sum_{i=1}^m \sigma_{A_i}(\theta))}$, with equality if there can be more arrivals than services in all states.}

	\subsection{Analysis of a two-server tandem network}
	\label{sec:analysis}
	In this paragraph, we compute new bounds of the violation probability for the backlog and delay for tandem networks on the small, yet representative example of Figure~\ref{fig:tandem}(a).

	\newcommand{\defi}{{\scriptstype \mathbf{(Def)}}}
	\newcommand{\theo}[1]{{\scriptstyle \text{\textbf{(Th.~#1)}}}}
	
	\subsubsection{Backlog violation probability}
	\label{ssec:backlog}
	Consider the network of Figure~\ref{fig:tandem}(a). From Theorems~\ref{th:bounds} and~\ref{th:e2e-sc}, the violation probability for backlog $b$ is 
	$$\p(q(t_3) \geq b) \leq \p(\exists t_1\leq t_2\leq t_3,~A_1(t_1, t_3) - S_1(t_1, t_2) - S_2(t_2, t_3) \geq b).$$
	The next computation is done in several steps: first in \eqref{eq:exbkunion} we partially use the union bound and sum on all $t_2$, then in~\eqref{eq:exbkmart},  we apply Theorem~\ref{th:martingale} with $Y = b + S_2(t_2, t_3)$, $u_0 = u_1 = t_2$, $v_0 = t_2$ and $v_1 = t_3$. This is valid for all $\theta \in[0, \theta_1^*]$. In~\eqref{eq:sr}, we use the $(\sigma(\theta),\rho(\theta))$ representations to bound the expectation, and finally, in~\eqref{eq:exbksum}, we sum all the terms. The sum is finite for all $\theta\in[0,\theta^*_2)$. For all $\theta \in[0,  \theta^*_1] \cap[0, \theta^*_2)$, 
	\begin{align}
		\p(q(t_3) \geq  b) & \leq \p\big(\sup_{t_1\leq t_2\leq t_3} A_1(t_1, t_3) - S_1(t_1, t_2) - S_2(t_2, t_3) \geq b\big) \notag\\ 
		 \ub& \leq \sum_{t_2\leq t_3} \p\big(\sup_{t_1\leq t_2} A_1(t_1, t_3) - S_1(t_1, t_2)  \geq b + S_2(t_2, t_3)\big) \label{eq:exbkunion} \\ 
		& \leq \sum_{t_2\leq t_3} \p\big(\sup_{\tau\geq 0} A_1(t_2-\tau, t_3) - S_1(t_2 - \tau, t_2)  \geq b + S_2(t_2, t_3)\big) \notag \\ 
		\theo{\ref{th:martingale}} &\leq \sum_{t_2\leq t_3} \xi_{A_1, S_1}(\theta)  e^{\theta \rho_{A_1}(\theta)(t_3-t_2)} \E\big[e^{-\theta (b + S_2(t_2, t_3))}\big]   \label{eq:exbkmart}\\ 
		 {\scriptstyle (\sigma, \rho)}& \leq \sum_{t_2\leq t_3} \xi_{A_1, S_1}(\theta) e^{\theta \rho_{A_1}(\theta)(t_3 - t_2) - \rho_{S_2}(\theta)(t_3-t_2) + \sigma_{S_2}(\theta) - b)}  \label{eq:sr}\\
		 \gs &\leq \frac{\xi_{A_1, S_1}(\theta) e^{\theta (\sigma_{S_2}(\theta) - b)}}{1 - e^{-\theta(\rho_{S_2}(\theta) - \rho_{A_1}(\theta))}}. \label{eq:exbksum}
	\end{align}
	One can first remark that with slightly modified constant terms, this formula is similar to the backlog violation probability for the network made only of server~2 and flow 1 (see Equation~\eqref{eq:vp-single-server-backlog}). 
	
	This formula can also be compared with the one obtained with the PMOO analysis from~\cite{BNS22a}: $\forall \theta \in [0, \min(\theta^*_1, \theta^*_2))$,  
	$$ \p(q(t) \geq b) \leq  \frac{e^{\theta (\sigma_{S_2}(\theta) + \sigma_{S_1}(\theta) + \sigma_{A_1}(\theta) - b)}}{\big(1 - e^{-\theta(\rho_{S_1}(\theta) - \rho_{A_1}(\theta))}\big)\big(1 - e^{-\theta(\rho_{S_2}(\theta) - \rho_{A_1}(\theta))}\big)}. $$
	
	Intuitively, if $\theta^*_1 < \theta^*_2$ (server 1 is the bottleneck), the optimal value taken for the PMOO formula is around $\theta^*_1$, and the value of $1 - e^{-\theta(\rho_{S_1}(\theta) - \rho_A(\theta))}$ be very small. The localized use of the martingale analysis then drastically improves the bound. On the contrary, if $\theta^*_2\leq\theta^*_1$, the gain is more limited as the optimal value for $\theta$ will approach $\theta^*_2$, and in the two expressions, the factor $(1 - e^{-\theta(\rho_{S_2}(\theta) - \rho_A(\theta))})$ is small resulting in a large prefactor in both analyses. The gain for the partial use of the martingale is then approximately $(1- e^{-\theta(\rho_{S_1}(\theta) - \rho_{A_1}(\theta))})^{-1}$ only. 
	
	The gain would then be much larger if the application of the martingale analysis could be localized at server 2. Unfortunately, this seems to be a much more difficult problem. Indeed, a first attempt for this would be to compute 
	\begin{align*}
		\p(q(t_3) \geq  b)  
		& \leq \sum_{t_1\leq t_3} \p\big(\sup_{t_1\leq t_2 \leq t_3} A_1(t_1, t_3) - S_1(t_1, t_2) -S_2(t_2, t_3) \geq b\big).
	\end{align*}
	In this latter expression, the process $(S_1(t_1, t_2) + S_2(t_2, t_3))_{t_2 \in [t_1, t_3]}$ is not a martingale, and the presented approach cannot be applied. 
	
	One solution is to assume that the first server is a constant-rate server: there exists a constant $C_1$ such that for all $(t_1, t_2)\in\N_{\leq}^2$, $S_1(t_1, t_2) = C_1(t_2-t_1)$. Then instead of fixing  $t_2$ for applying Theorem~\ref{th:martingale} at server $2$, one can fix $k = t_2-t_1$. In line~\eqref{eq:repl}, $t_1$ is replaced by $t_2 - k$. In line~\eqref{eq:const}, we 
	use the constant rate service property: $S_1(t_2 - k, t_2) = C_1k$ does not depend on $t_2$. Theorem~\ref{th:martingale} is then applied in line~\eqref{eq:appl} with $u_1 = t_3-k$, and $u_0 = v_0 = v_1 = t_3$.
	\begin{align}
		\p(q(t_3) \geq b) & \leq \p\big(\sup_{t_1\leq t_2\leq t_3} A_1(t_1, t_3) - S_1(t_1, t_2) - S_2(t_2, t_3) \geq b\big) \notag \\ 
		\ub\label{eq:repl} & \leq \sum_{ k \geq 0} \p\big(\sup_{t_2\leq t_3} A_1(t_2-k , t_3) - S_1(t_2-k, t_2) - S_2(t_2, t_3) \geq b\big) \\
		& \leq \sum_{k \geq 0} \p\big(\sup_{t_2\leq t_3} A_1(t_2-k , t_3) - S_2(t_2, t_3) \geq b + C_1k\big) \label{eq:const}\\
		& \leq \sum_{k \geq 0} \p\big(\sup_{\tau \geq 0} A_1(t_3-k - \tau , t_3) - S_2(t_3 - \tau, t_3) \geq b + C_1k\big) \notag\\
		\theo{\ref{th:martingale}} &\leq \sum_{k \geq 0} \xi_{A_1, S_2}(\theta)  e^{-\theta (b - \rho_{A_1}(\theta)k)} e^{-\theta C_1 k}  \label{eq:appl} \\ 
		\gs&\leq \frac{\xi_{A_1, S_2}(\theta) e^{-\theta  b}}{1 - e^{-\theta(\rho_{C_1} - \rho_{A_1}(\theta))}}.  \notag 
	\end{align}
	Remark that if $S_2$ is also a constant rate-server, and end-to-end server is $S_1\land S_2$, and the network can be analyzed as a single-server network. 
	
	\newcommand{\cp}{\text{{\scriptstyle \textbf{(Eq.~(\ref{eq:chauchy_n}))}}}}
	
	\subsubsection{Delay violation probability}
	\label{ssec:delay}
	Let us now focus on the computation of the violation probability of the delay, again with the network of Figure~\ref{fig:tandem}(a). 
	Recall that we have, from Theorems~\ref{th:bounds} and~\ref{th:e2e-sc},
	 {\small \begin{align*}
		d(t_3\!-\!T\!+\!1) \geq T & \Rightarrow \exists t_1\leq t_3-T,~A_1(t_1, t_3\!-\!T\!+\!1) > \inf_{t_1\leq t_2\leq t_3} S_1(t_1, t_2) + S_2(t_2, t_3)\\
		&\Rightarrow \exists t_2 \leq t_3,~\sup_{t_1\leq t_2\land t_3-T} A_1(t_1, t_3\!-\!T\!+\!1) - S_1(t_1, t_2) > S_2(t_2, t_3).
	\end{align*}}
	One can then write, using the union bound,
	{\small \begin{align*}
		\p(d(t_3\!-\!T\!+\!1) \geq T) & \leq \sum_{t_2\leq t_3} \p\big(\!\!\sup_{t_1\leq  t_2\land t_3-T}\!\! A_1(t_1,t_3\!-\!T\!+\!1) - S_1(t_1, t_2) > S_2(t_2, t_3)\big)\\ 
		&=  \sum_{t_2\leq t_3-T} \p\big(\sup_{t_1\leq  t_2} A_1(t_1, t_3\!-\!T\!+\!1) - S_1(t_1, t_2) > S_2(t_2, t_3)\big)\\ &  \hspace{-1cm}+  \sum_{t_3-T < t_2\leq t_3} \p\big(\sup_{t_1\leq  t_3-T} A_1(t_1, t_3\!-\!T\!+\!1) - S_1(t_1, t_2) > S_2(t_2, t_3)\big).
	\end{align*}}
	
	In the last equality, we distinguish two cases, depending on how $t_2$ and $t_3-T$ compare. We will deal with them separately. 
	
	In the first sum sign, that we denote $P_1$, one can apply Theorem~\ref{th:martingale} with $Y = S_2(t_2, t_3)$, $u_0 = u_1 = v_0 = t_2$ and $v_1 = t_3-T+1$. For all $\theta_1 \in [0, \theta_1^*] \cap [0, \theta_2^*)$, 
	{\small \begin{align*}
		P_1 & \leq  \sum_{t_2\leq t_3-T}\xi_{A_1, S_1}(\theta_1) e^{\theta_1\sigma_{S_2}(\theta_1)}e^{-\theta_1(\rho_{S_2}(\theta_1)(t_3-t_2)-\rho_{A_1}(\theta_1)(t_3-T +1-t_2))}\\ 
		\gs & \leq  \frac{\xi_{A_1, S_1}(\theta_1) e^{\theta_1(\sigma_{S_2}(\theta_1) + \rho_{A_1}(\theta_1)-\rho_{S_2}(\theta_1)T)}}{1-e^{-\theta_1(\rho_{S_2}(\theta_1) - \rho_{A_1}(\theta_1))}}.
	\end{align*}}
	With slightly modified constant terms, one can recognize the delay violation probability for the network made only of server 2 and flow 1.

	In the second sum sign, that we denote $P_2$, let us apply Theorem~\ref{th:martingale} with $u_0 = u_1 = t_3-T$, $v_1 = t_3-T + 1$ and $v_0 = t_2$. We then obtain for all $\theta_2 \in [0, \theta_1^*] \cap [0, \theta_2^*)$, 
	\begin{align*}
		P_2 & \leq \sum_{t_3-T < t_2 \leq t_3} \xi_{A_1, S_1}(\theta_2) e^{\theta_2\sigma_{S_2}(\theta_2)}e^{-\theta_2(\rho_{S_1}(\theta_2)(t_2-t_3+T)-\rho_{A_1}(\theta_2)+\rho_{S_2}(\theta_2)(t_3-t_2))} \\
		& \leq \xi_{A_1, S_1}(\theta_2)e^{\theta_2(\sigma_{S_2}(\theta_2)  + \rho_{A_1}(\theta_2) - \rho_{S_1}(\theta_2))}  \sum_{u= 0}^{T-1}  e^{-\theta_2(\rho_{S_1}(\theta_2)(T - 1 -u)+\rho_{S_2}(\theta_2)(u))}\\
		 & \leq  [z^{T-1}] \frac{\xi_{A_1, S_1}(\theta_2)e^{\theta(\sigma_{S_2}(\theta_2)  + \rho_{A_1}(\theta_2) - \rho_{S_1}(\theta_2))}}{\big(1-e^{-\theta_2\rho_{S_1}(\theta_2)}z\big)\big(1-e^{-\theta_2\rho_{S_2}(\theta_2)}z\big)}.
	\end{align*}
	In the last line, we recognize the $T-1$-th term of the Cauchy product~in Equation~\eqref{eq:chauchy_n},
	and with slightly modified constant term, this is the  $T-1$-th term of the service bounding generating series of the end-to-end server for flow 1. 
	Finally, the delay violation probability can be bounded by 
	\begin{multline*}
		\p(d(t_3-T) \geq  T) \leq   \frac{\xi_{A_1, S_1}(\theta_1) e^{\theta_1(\sigma_{S_2}(\theta_1) + \rho_{A_1}(\theta_1)-\rho_{S_2}(\theta_1)T)}}{1-e^{-\theta_1(\rho_{S_2}(\theta_1) - \rho_{A_1}(\theta_1))}} \\  +  [z^{T-1}] \frac{\xi_{A_1, S_1}(\theta_2)e^{\theta_2(\sigma_{S_2}(\theta_2) + \rho_{A_1}(\theta_2) - \rho_{S_1}(\theta_2))}}{\big(1-e^{-\theta_2\rho_{S_1}(\theta_2)}z\big)\big(1-e^{-\theta_2\rho_{S_2}(\theta_2)}z\big) },
	\end{multline*}
	and to minimize this bound, $\theta_1$ and $\theta_2$ can be optimized independently. 
	
	This computation suffers from the same limitation as for the backlog: it has not been possible yet to apply Theorem~\ref{th:martingale} to the second server directly, unless assuming that the first server is a constant-rate server.

	\subsection{Main Theorem for Tandem Networks}
	\label{ssec:general-result}
	In this section, we generalize the computations made for the two-server tandem. Following the approach, we apply the martingale analysis (and the Doob's inequality) locally at one server, and the union and Chernoff bounds for the other servers. 
	From the discussion when computing the backlog in the two-server tandem, the martingale analysis cannot be applied to any server $h$ , and some conditions have to be fulfilled by server $h$. Let us first give these conditions.

	\begin{assumption}[Conditions for martingale analysis at server $h$] Server $h$ must satisfy 
		\begin{itemize}
			\item[$(H_6)$] for all $j< h$, server $j$ is a constant-rate server, {\em i.e.,} there exists $C_j$ such that $S_j(u, t) = C_j(t-u)$;
			\item[$(H_7)$]  for all flow $i\in\N_m,~f_i \leq h \implies \ell_i\geq h$. In other words, no flow arrived before server $j$ departs before server $h$. 
		\end{itemize}
	\end{assumption}
	
	Assumption $(H_6)$ is a direct consequence of the discussion for the two-server backlog: the martingale part of the analysis could be applied to the second server only if the first server was a constant-rate server. Here, $(H_6)$ assumes  that the all upsteam servers of server $h$ are constant-rate servers. 
	
\begin{figure}[htbp]
		\centering
		
		\begin{tikzpicture}[server/.style={shape=rectangle,draw,minimum height=.8cm,inner xsep=3ex}]
			\node[server,name=S1] at (0,0) {$1$};
			\node[server,name=S2] at (2,0) {$3$};
			\draw[->, thick, red] (-1,0) node[left] {$1$} -- (3,0);
			\draw[->, thick, blue] (-1,0.3) node[left] {$2$} to[out=-10,in=-170] (0.7,.3);
			\draw[->, thick, green!70!black]  (1,-0.3) node[left] {$3$} to[out=10,in=170] (3,-.3);
		\end{tikzpicture}
		\caption{Network obtained from the network in Figure~\ref{fig:tandem}(b) after removal of server 2.}
		\label{fig:remove}
	\end{figure}
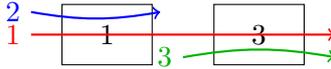

	Assumption $(H_7)$ is also related to this. Assume for example that flow 2 crosses server 1 only (as in Figure~\ref{fig:remove} with only flows 1 and 2). For the viewpoint of flow 1, server 1 offers the service $S_1 - A_2$, which is not constant-rate anymore (unless $A_2$ is deterministic, but we focus on stochastic arrival processes in this paper). As a consequence, the martingale part of the analysis could not be applied to server 2. Now, if flow 2 also crossed server 2, flow 2 could be incorporated in the partial analysis.

	\paragraph{Transforming a network by removing one server}
	We also saw in the previous paragraph that the violation probability bounds look like the violation probability bound of a network reduced to the second server only. This is also the case for general tandem networks, when terms will be alike the violation probability of a tandem with one server less, as if the server where the martingale analysis is performed were removed. 
	{The goal of this transformation is to separate the part where the usual MGF-SNF analysis with union bound is used (tandem without server $h$) and where the Doob's inequality is applied (server $h$). The tandem network describe next represent the part of the network that will be analyzed with the usual MGF-SNC method.}
	
	Consider a tandem network described by the notations in $(H_1)$ and $(H_2)$. The tandem network obtained by removing server $h$ is constructed as follows: 
	\begin{itemize}
		\item it is made of servers $1,\ldots, h-1, h+1\ldots, n$;
		\item for all flow $i$, its  path of flow $i$ is unchanged unless it crosses server $h$, in case it goes directly from server $h-1$ to server $h+1$;
		\item flows originally crossing server $h$ only are removed;
		\item arrival processes and service processes of the remaining flows and servers are unchanged. 
	\end{itemize}
	
	For example, the network obtained from the one of Figure~\ref{fig:tandem}(b) by removing server 2 is depicted in Figure~\ref{fig:remove}. The notation $S_{e2e}^{(-h)}$ refers to the end-to-end server where server $h$ has been removed. 
	
	We can now state our main result, proved in Paragraph~\ref{app:main}.

	\begin{theorem}
		\label{th:main}
		Consider a tandem network satisfying notations and assumptions given in $(H_1)$--$(H_7)$, and $\Theta = [0, \theta^*_h] \cap [0, \inf_{j\neq h} \theta^*_j)$. For all $\theta, \theta_1 \in \Theta$ and $\theta_2 \in[0, \theta^*_h]$, 
		\begin{enumerate}
			\item the violation probability of the backlog for flow 1 satisfies
			$$\p(q(t) \geq b) \leq\frac{\xi_{(A_i)_{i\in \fl(h)}, S_h}(\theta) e^{-\theta \sigma_{A_1}(\theta)}}{e^{\theta( \sum_{i\in \fl(h) - H} \sigma_{A_i}(\theta))}} F_{S_{e2e}^{(-h)}}(\theta, e^{\theta \rho_{A_1}(\theta)}) e^{-\theta b},$$
			\item a delay bounding generating series for the delay of flow 1 is
			\begin{multline*}\frac{\xi_{(A_i)_{i\in \fl(h)}, S_h}(\theta_1)}{e^{\theta_1 \sum_{i\in\fl(h) \setminus H} \sigma_{A_i}(\theta_1)}} F_{d, S_e2e^{(-h)}}(\theta_1, z) \\  + \frac{\xi_{(A_i)_{i\in \fl(h)}, S}(\theta_2)  e^{-\theta_2(\rho_{S_h}(\theta_2) - \sum_{i\in\fl(h)} \rho_{A_i}(\theta_2))}}{e^{\theta_2(\sigma_{S_h}(\theta_2) + \sum_{i\in\fl(h)\setminus\{1\}}\sigma_{A_i}(\theta_2))}} z F_{S_{e2e}}(\theta_2, z),
			\end{multline*}
		\end{enumerate} 
		where $H$ is the set of flows crossing only server $h$, and the bounding generating functions are computed as in Theorems~\ref{th:bounds} and~\ref{th:e2e-bg}. 
	\end{theorem}
	
	The violation probability of the backlog is similar to the one with MGF-based SNC obtained by combining Theorems~\ref{th:bounds} and~\ref{th:e2e-bg} when server $h$ is removed. This can be explained as follows: the end-to-end service bounding generating function $F_{S_{e2e}}(\theta, z)$ in Theorem~\ref{th:e2e-bg} is the product of $n$ geometric functions, each representing one server. The product of generating function represents the Cauchy product of series, and in our case, corresponds to the union bound at each server. When martingale analysis is used locally at one server, then there is no union bound for that server, but the union bound is still used for the other servers and $F_{S_{e2e}}^{(-h)}(\theta, z)$ naturally appears. The prefactor is some reorganization of the terms, and compensating the $e^{\theta\sigma_{A_i}(\theta)}$ that are already taken into account in a different way in $\xi_{(A_i)_{i\in \fl(h)}, S_h}(\theta)$. {The pre-factor $\frac{\xi_{(A_i)_{i\in \fl(h)}, S_h}(\theta) e^{-\theta \sigma_{A_1}(\theta)}}{e^{\theta( \sum_{i\in \fl(h) - H} \sigma_{A_i}(\theta))}} \leq e^{\theta (\sigma_{S_h}(\theta) + \sum_{i\in H} \sigma_{A_i}(\theta))}$, which are terms that would appear in the MGF-SNC analysis of the complete end-to-end tandem, but not when server $h$ is removed.}
	
	The violation probability of the delay has two parts. The interpretation of the first term  is similar to the one of the violation probability of the backlog. The second term comes from the case distinction that was done. In the second case, the sum has a finite number of terms, and  the factor $e^{-\theta \rho_{A_1}(\theta)}$ does not appear with a power of some time variables. This means that this term will be similar to the end-to-end service.

	\section{An Alternative Analysis with Martingales}
	\label{sec:AK}
	Another way of using martingales was attempted in the literature, first in~\cite{AK11}, and recently used in~\cite{YCL22}. In this section we explain why this approach is not sound and should not be used before being corrected. The authors study a tandem network in the absence of cross traffic ($m=1$ and flow 1 crosses all the servers), and where each server $j$ is a work-conserving server with i.i.d. increments. 
	
	Let us detail the approach used in~\cite{AK11} and~\cite{YCL22}. Here we adapt it to the computation of the  backlog violation probability instead of the  delay. The aim is to simplify the notations,  the calculations remain the same. The first step of the analysis is:
	\begin{align}
		\p(q(t_{n+1}) \geq b)& \leq \p\big(\sup_{t_1\leq \cdots \leq t_{n+1}} A(t_1, t_{n+1}) - \sum_{j=1}^n S_j(t_j, t_{j+1})\big) \notag\\ 
		&\leq \p\big(\sup_{t_1\leq \cdots \leq t_{n+1}} [A(t_1, t_{n+1}) - \rho_A(\theta^*)(t_{n+1} - t_1)]  \notag\\
		& \hspace{1cm} -  \sum_{j=1}^n [S_j(t_j, t_{j+1}) - \rho_{S_j}(\theta^*)(t_{j+1} - t_j)] \geq b\big)  \label{eq:ak1}\\ 
		&\leq \p\big([\sup_{t_1 \leq t_{n+1}} A(t_1, t_{n+1}) - \rho_A(\theta^*)(t_{n+1} - t_1)] \notag\\
		& \hspace{1cm}  + \sum_{j=1}^{n-1} [\sup_{t_{j} \leq t_{j+1} \leq t_{n+1}}\rho_{S_j}(\theta^*)(t_{j+1} - t_j) -  S_j(t_j, t_{j+1}) ] \notag \\
		& \hspace{1cm}  + [\sup_{t_n\leq t_{n+1}}  \rho_{S_n}(\theta^*)(t_{n+1} - t_n) - S_j(t_j, t_{j+1}) ] \geq b\big). \label{eq:ak2}
	\end{align}
	In Inequality~\eqref{eq:ak1}, the authors use that $\rho_A(\theta^*) \leq \rho_{S_j}(\theta^*)$ for all $j\in\{1, \ldots, n\}$. In Inequality~\eqref{eq:ak2}, they separate the global supremum into $n$ independent suprema, one per server. This is a step where the accuracy of the probability bound is lost, and the aim of the next step is to give tight bounds on the violation probabilities of each supremum. More precisely, the second step is bounding tightly, for all $j<n$, 
	$$P_j = \p\big(\sup_{t_j\leq t_{j+1}\leq t_{n+1}} \rho_{S_j}(\theta^*)(t_{j+1}- t_j) - S_j(t_j, t_{j+1})  \geq x\big).$$
	
	Let us consider the quantities: 
	\begin{itemize}
		\item $M_j(t_j, t_{j+1}) = e^{-\theta^*(\rho_{S_j}(\theta^*)(t_{j+1} - t_j) - S_j(t_j, t_{j+1}))}.$ From Lemma~\ref{lem:duffield},  $(M_j(t_{j+1} - \tau, t_{j+1})_{\tau\geq 0}$ is a martingale with respect to the filtration $\cF_{S_j}(t_{j+1} - \tau, t_{j+1})$;
		\item $\widetilde{M}_j(t_{j+1})	= \sup_{0\leq t_{j} \leq t_{j+1}} M_j(t_j, t_{j+1})$. The authors in~\cite{AK11, YCL22} show that $\widetilde{M}_j(t_{j+1})_{t_{j+1}\geq 0}$ is a sub-martingale with respect to the filtration $\cF_{S_j}(0, t_{j+1})$.
	\end{itemize}
	
	They then claim that a bound for $P_j$ is 
	\begin{align}
		P_j & =	\p\big(\sup_{0\leq t_j\leq t_{j+1}\leq t_{n+1}} S_j(t_j, t_{j+1}) - \rho_{S_j}(\theta^*)(t_{j+1}- t_j) \geq x\big) \notag \\
		& = \p\big(\sup_{0\leq t_j\leq t_{j+1}\leq t_{n+1}} M_j(t_j, t_{j+1}) \geq e^{\theta^* x}\big) \notag \\
		& = \p\big(\sup_{0\leq t_{j+1} \leq t_{n+1}} \widetilde{M}_j(t_{j+1}) \geq e^{\theta^* x}\big) \notag \\
		& \leq \E[ \widetilde{M}_j(t_{n+1})] e^{-\theta^* x}  \label{eq:sub-mart} \\ 
		& \leq e\E[M_j(0, t_{n+1})]e^{-\theta^* x} \label{eq:rao}\\ & =  e e^{-\theta^* x}. \label{eq:exp1} 
	\end{align}
	Inequality~\eqref{eq:sub-mart} is the direct application of Doob's inequality for sub-martingales. Equality~\eqref{eq:exp1} is because $\E[M_j(0, t_{n+1})] = 1$ by construction of the martingale.

	To prove Inequality~\eqref{eq:rao}, the authors refer to Theorem 3.7 by Rao~\cite{Rao07} stating that if $(X_t)_{t\in\N}$ is a demi-submartingale, then for all $r>0$, $\E[e^{r\max_{0\leq u\leq t} X_u}] \leq e\E[e^{rX_t}]$.

	The application of this inequality would be on the submartingale $\widetilde{M}_j(t_{j+1})$, and one would get 
	$$\E\big[e^{r \widetilde{M}_j(t_{n+1})}\big]  \leq e \E\big[e^{r e^{-\theta^*(S_j(0, t_{n+1}) - \rho_{S_j}(\theta^*)(t_{n+1}))}}\big],$$
	which is not the desired inequality. 
	
	To obtain $\E[\sup_{0\leq t_{j} \leq t_{n+1}} M_j(\theta^*,t_{j}, t_{n+1})] \leq e\E[ M_j(\theta^*, 0, t_{n+1})]$, one would require that $(\sup_{0\leq t_j\leq t_{n+1}} \theta^*(\rho_{S_j}(\theta^*)(t_{n+1} - t_j) - S_j(t_j, t_{n+1})))_{t_{n+1}\geq 0}$ is a sub-martingale. 
	
	Let us define $N(t) = \sup_{0\leq u\leq t} \rho_{S_j}(\theta^*)(t-u) - S_j(u, t)$. First, with $s_t = S(t, t+1)$, we have 
	\begin{align*}
		\E[N(t+1)| N(t)] & = \E[\sup_{0\leq u\leq t+1 } \rho_{S_j}(\theta^*)(t+1 - u) - S_j(u, t+1)~|~N(t)] \\
		& = \E[\sup_{0\leq u\leq t } \rho_{S_j}(\theta^*)(t - u) - S_j(u, t) + \rho_{S_j}(\theta) -  s_t\lor 0 ~|~N(t)]\\
		& = \E[N(t) +  \rho_{S_j}(\theta) - s_t \lor 0~|~N(t)].
	\end{align*}
	But from Jensen's inequality, since $f: x\mapsto e^{-\theta^* x}$ is convex, $\E[e^{-\theta^* s_t}] = e^{-\theta^*\rho_{s_j}(\theta^*)} \geq e^{-\theta^*\E[s_t]}$ and $\E[s_t] \geq \rho_{s_j}(\theta^*)$.  
	So we cannot conclude that $N(t)$ is a sub-martingale (without the maximum with 0, one could easily conclude that $(N(t))_{t\geq 0}$ is a super-martingale). The proof is then incomplete the result might be erroneous.
	
\begin{figure}[htbp]
		\centering
		\includegraphics[scale=0.5]{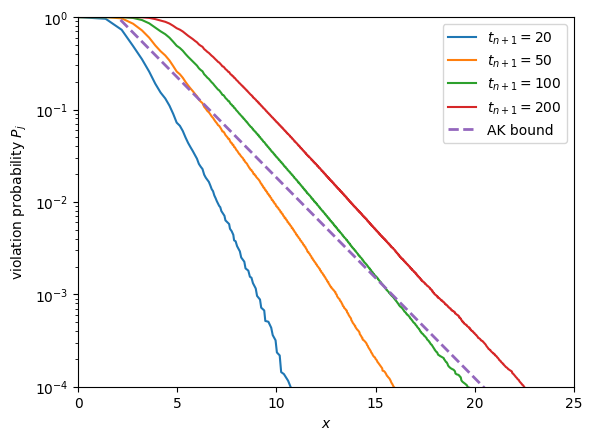}
		\caption{Illustration of the erroneous bound for $P_j$ in \cite{AK11}.}
		\label{fig:ak}
	\end{figure}

	Figure~\ref{fig:ak} illustrates the erroneous bound. We choose for $s_j$ the Poisson distribution with parameter $1$ and $\theta^* = 0.5$. On the one hand, we depict in solid lines the violation probability $P_j$ in function of $x$ for different values of $t_{n+1}$. We observe that $P_j$ strongly depends on the time horizon $t_{n+1}$. On the other hand, we depict the bound $ee^{-\theta^*x}$ in dashed line. For a time horizon larger than 50, the bound is violated for small values. For $t_{n+1} = 200$, the bound is almost always violated. The observation is also in line with Kolmogorov's zero-one law: if a tail event can happen with positive probability in an i.i.d sequence, it will hapen infinitely often. 
	
	\section{Numerical Evaluation}
	\label{sec:numerical}
	In this section, we compare the bounds obtained with simulation and the state of the art. 
	It has been demonstrated in~\cite{BNS22a} that the PMOO analysis outperforms by far the other SNC methods from the state-of-the-art. As a consequence, we only compare against this method. 
	We perform two different types of experiment: first we compare the violation probability for different values of target delays. For this type of experiments, we run 10 independent simulations with $10^8$ time steps. In the second type of experiment, we use a free parameter (transmission probability or capacity of a server) and observe the quality of the bound for a violation probability of $10^{-4}$. For this type of experiment, we run one simulation with $10^7$ time steps. 
	
	We also chose to only compare the end-to-end delays for the networks. The results for the backlog bounds show similar comparisons. 
	
	\subsection{Two-server case}
	Let us first consider the network of Figure~\ref{fig:tandem}(a). In this experiment, we assume a Markov modulated on-off process, that is an MMP with two states, \on~and \off. In the \off~state, there is no arrival and in the \on~states, arrivals arrive according to a given distribution. Here, let us take $P_{\off, \on} = 0.7$ and $P_{\on, \off} = 0.1$, and in the \on~state, the arrivals follow a Poisson distribution of intensity 2. 
	We assume that server 1 serves 5 packets with probability $p$, and 0 packet otherwise, and server 2 serves 6 packets with probability $q$ and 0 otherwise. 
	
\begin{figure}[htbp]
		\centering
		\includegraphics[scale=0.4]{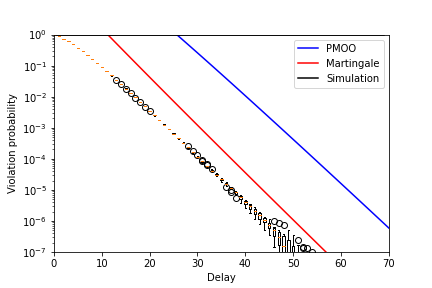}~%
	\includegraphics[scale=0.4]{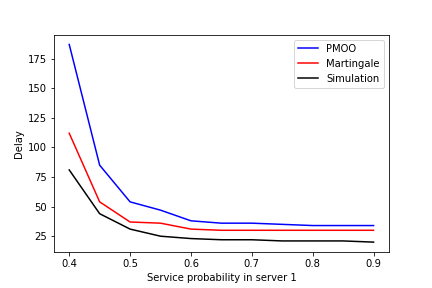}
		\caption{Two-server tandem network with Bernoulli service process. (left) violation probability in function of the target delay; (right) delay bound in in function of the service probability of server 1.}
		\label{fig:two-server-delay}
	\end{figure}

	Figure~\ref{fig:two-server-delay}(left) compares the violation probability in function of a target delay obtained by different methods: simulation, PMOO and with our martingale bound, when $p=q=0.5$. One can see that our new methods improves the PMOO analysis. For example, with a violation probability of $10^{-4}$, the simulation delay is 27, the delay obtained with PMOO is 54 and with our method 37. The gap is then reduced by 63\%.
	
	We now study the improvement obtained when $p$ is varying. We still fix $q=0.5$.  Figure~\ref{fig:two-server-delay}(right) shows the delays obtained for a fixed violation probability of $10^{-4}$ with the different methods. When $p$ is large, server 2 is the bottleneck. Hence, the improvement of the martingale analysis compared to PMOO is limited. On the contrary, when server 1 is the bottleneck (small values of $p$), the improvement becomes larger, specially when the network is heavily loaded. 
	
	\subsection{Interleaved tandem network}
\begin{figure}[htbp]
		\centering
		\includegraphics[scale=0.4]{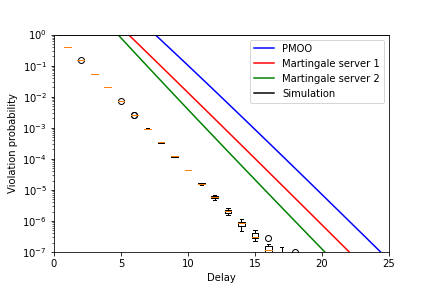}~%
	\includegraphics[scale=0.4]{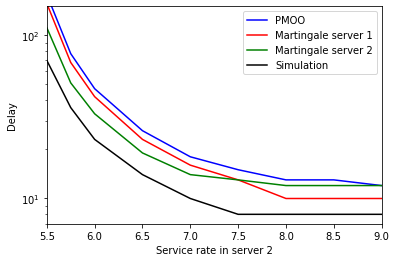}
		\caption{Interleaved tandem network with constant-rate service. (left) violation probability in function of the target delay; (right) delay bound in function of the capacity of the second server.}
		\label{fig:num-interleaved-delay}
	\end{figure}

	Let us now focus on the network of Figure~\ref{fig:tandem}(right) and assume that all servers are constant-rate.  Theorem~\ref{th:main} can be applied for $h\in\{1, 2\}$. We assume that all flows share the same characteristics as in the previous example. 
	In Figure~\ref{fig:num-interleaved-delay}(left), we set the rates of the server as $C_1 = 5$, $C_2 = 7$ and $C_3=6$. The bottleneck is the server 2. The figure compares simulation, PMOO and the bound obtained by Theorem~\ref{th:main} applied respectively to server 1 and 2. One can observe that, as expected, the gain is more important when applying the martingale analysis to server 2. 
	
	In Figure~\ref{fig:num-interleaved-delay}(right), $C_2$ is varying from 5.5 to 9, so depending on its value, server 1 or server 2 is the bottleneck. We observe that when server 2 is the bottleneck ($C_2 \leq 7.5$) then the better solution is to apply the martingale analysis at server 2 and when server 1 is the bottleneck ($C_2 \geq  7.5$), the better solution is to apply the martingale at server 1. The reduction of the pessimism gap compared to PMOO ranges from 33\% to 75\%. The reduction is the smallest for $C = 7.5$, when servers 1 and 2 are bottlenecks.
	
	\subsection{Sink-trees tandem network}
	\label{ssec:sink-tree}
	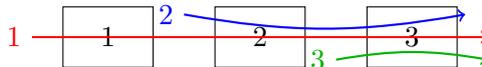
\begin{figure}[htbp]
		\centering
		\begin{tikzpicture}[server/.style={shape=rectangle,draw,minimum height=.8cm,inner xsep=3ex}]
			\node[server,name=S1] at (0,0) {$1$};
			\node[server,name=S2] at (2,0) {$2$};
			\node[server,name=S3] at (4,0) {$3$};
			\draw[->, thick, red] (-1,0) node[left] {$1$} -- (5,0);
			\draw[->, thick, blue] (1,0.3) node[left] {$2$} to[out=-10,in=-170] (4.7,.3);
			\draw[->, thick, green!70!black]  (3,-0.3) node[left] {$3$} to[out=10,in=170] (5,-.3);
		\end{tikzpicture}
		\caption{Example of sink-tree tandem network.}
		\label{fig:sink-tree}
	\end{figure}
\begin{figure}[htbp]
		\centering
		\includegraphics[scale=0.4]{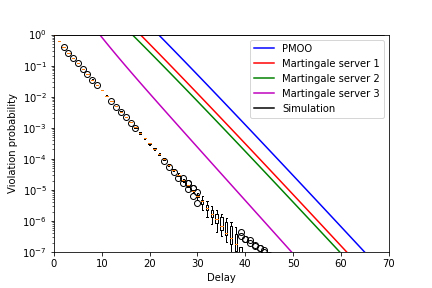}~%
		\hspace{0.5cm}
		\includegraphics[scale=0.4]{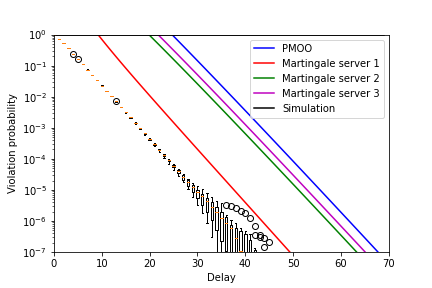}
		\caption{Sink-tree tandem network. Violation probability in function of the target delay  when server 3 (left) or  server 1 (right) is the bottleneck.}
		\label{fig:num-sinktree-delay}
	\end{figure}

	A sink-tree tandem network is a tandem network for which all flows end at the last server. For example, Figure~\ref{fig:sink-tree} is a sink-tree tandem. Because all flows end at the last server, assumption $(H_7)$ is always satisfied, and if we further assume that all servers are constant rate servers, then Theorem~\ref{th:main} can be applied at each server. Let us focus on the sink-tree of Figure~\ref{fig:sink-tree}. The arrival processes are the same as in the previous examples. In Figure~\ref{fig:num-sinktree-delay}(left), the service rates are $C_i = 3 + i$, $i\in\{1, 2, 3\}$. In that case, $\theta^*_3 < \theta^*_2 < \theta^*_1$ and server 3 is the bottleneck. One can check that the best option is to apply the martingale analysis to server 3. While using the martingale always improved the performance bounds compared to PMOO, one can notice that applying it on server 1 or 2 only marginally improves the bounds, the pessimism gap is reduces by more than 50\% compares to PMOO with the martingale at server 3.
	
	In Figure~\ref{fig:num-sinktree-delay}(right), the service rates are $C_i = 3 i -1$, $i\in\{1, 2, 3\}$, and $\theta^*_1 < \theta^*_2 < \theta^*_3$. Server 1 is then the bottleneck, and the results are reversed: applying the martingale at server 1 reduces the pessimism gap the most. 
	
	\section{Proofs}
	\label{sec:proofs}
	\subsection{Proof of Lemma~\ref{lem:duffield}}
	\label{sec:proofl1}
	To prove this lemma, we need to prove that $\E[M_A(\theta, u-1, v)~|~X(u)] = M_A(\theta, u, v)$. 
	{\small \begin{align}
		\E[M_A(\theta, u\!-\!1, v)|\cF_A(u, v)] & = \E[e^{\theta (A(u-1, v) - \rho_A(v-u+1))} 	\nu_A(\theta)_{X_A(u-1)}|\cF_A(u, v)] \notag \\ 
		& = e^{-\theta \rho_A(v-u+1)} \E[e^{\theta(a_{u\!-\!1} + A(u, v))}\nu_A(\theta)_{X_A(u\!-\!1)}|\cF_A(u, v)] \notag \\
		& = e^{-\theta \rho_A(v-u+1)} e^{\theta A(u, v)} \E[e^{\theta a_{u\!-\!1}} \nu_A(\theta)_{X_A(u\!-\!1)}|X(u)] \label{eq:duf1}\\
		& \hspace{-4cm} = e^{-\theta \rho_A(v-u+1)} e^{\theta A(u, v)} \sum_{x\in\cX_A} \E[e^{\theta a_{u-1}} \nu_A(\theta)_{x}|X(u), X(u\!-\!1) = x] \p(X(u\!-\!1) = x|X(u)) \label{eq:duf2}\\
		& = e^{-\theta \rho_A(v-u+1)} e^{\theta A(u, v)} \sum_{x\in\cX_A} \varphi_x(\theta) \nu_A(x) p^r_{X(u), x} \notag\\
		& = e^{-\theta \rho_A(v-u+1)} e^{\theta A(u, v)} \sum_{x\in\cX_A} \psi_A(\theta)_{X(u), x} \nu_a(\theta)_x \notag\\
		& = e^{-\theta \rho_A(v-u+1)} e^{\theta A(u, v)} e^{\theta\rho_A(\theta)}  \nu_a(\theta)_{X(u)} \label{eq:duf3}\\
		&= M_A(\theta, u, v). \notag 
	\end{align}}
	In line~\eqref{eq:duf1}, we remove from the expectation the terms that are completely defined with the filtration, and the rest of the terms depend on the filtration only with $X(u)$ (Markovian assumption). In line~\eqref{eq:duf2}, we decompose on all the possibilities on the state $X(u-1)$, and replace in the next lines by the notations defined in Paragraph~\ref{ssec:mmp}. In line~\eqref{eq:duf3}, we apply that $\nu_a(\theta)$ is a right eigenvector of  $\varphi_A(\theta)$ associated with eigenvalue $e^{\theta\rho_A}$, which  concludes the proof. 
	
	\subsection{Proof of Theorem~\ref{th:martingale}}
	\label{app:martingale}
	
	In the proof of Theorem~\ref{th:martingale}, we will need the two following lemmas.
	
	\begin{lemma}
		\label{lem:indep_chernoff}
		If $X$ and $Y$ are two independent random variables, and if there exist $K, \theta \in \R_+$ such that for all $b\in\N$, $\p(X \geq  b) \leq Ke^{-\theta b}$. Then $\p(X \geq  Y) \leq K \E[e^{-\theta Y}]$.
	\end{lemma}
\proof{
	\begin{align*}
					\p(X \geq  Y) & = \sum_{b\in \N} \p(X\geq b) \p(Y = b) \\ 
					&  \leq \sum_{b\in \N} Ke^{-\theta b} \p(Y = b) \\ 
					& = K\E[e^{-\theta Y}]. 
				\end{align*}
			}

	\begin{lemma}
		\label{lem:exp}
		Consider $A$ a process generated according to an MMP with exponential transition matrix $\psi(\theta)$, stationary distribution $\pi$, and other notations defined as in Paragraph~\ref{ssec:mmp}. Assume that $X(u)$ the state at time $u$ is distributed according to $\pi$.  Then for all $u\leq v$,
		$$\E[\nu(\theta)_{X(u)} e^{\theta A(u, v)}] =  e^{-\theta \rho_A(\theta)(u-v)}.$$
	\end{lemma}
	\proof{
		First, for all $\theta$ such that $\psi_A(\theta)$ is finite, we have for all $x_0\in\cX$
		\begin{align}
			\E[e^{\theta A(u, u+k)}~|~X(u) = x_0] & = \sum_{x_1, \ldots, x_{k}\in\cX} \prod_{\ell = 0}^{k-1} \varphi_{x_{\ell}}(\theta) P_{x_\ell, x_{\ell+1}} \notag \\
			& = \sum_{x_1, \ldots, x_{k}\in\cX} \prod_{\ell = 0}^{k-1} \varphi_{x_{\ell}}(\theta) \frac{\pi_{x_{\ell+1}}}{\pi_{x_{\ell}}} P^r_{x_{\ell+1}, x_{\ell}} \notag \\
			& = \sum_{x_1, \ldots, x_{k}\in\cX} \frac{\pi_{x_k}}{\pi_{x_0}} \prod_{\ell = 0}^{k-1} \varphi_{x_{\ell}}(\theta)  P^r_{x_{\ell+1}, x_{\ell}} \notag \\
			& = \sum_{x_k\in\cX} \frac{\pi_{x_k}}{\pi_{x_0}}  \sum_{x_1, \ldots, x_{k-1}\in\cX}\prod_{\ell = 0}^{k-1} \psi(\theta)_{x_{l+1}, x_\ell}\notag \\
			& = \sum_{x_k\in\cX} \frac{\pi_{x_k}}{\pi_{x_0}}  (\psi(\theta)^{k})_{x_k, x_0}. \label{eq:ksteps}
		\end{align}

		Then, 	
		\begin{align}
			\E[\nu_A(\theta)_{X(u)} e^{\theta A(u, v)}] & = \sum_{x\in \cX} \E[\nu_A(\theta)_x e^{\theta A(u, v)}~|~X(u) = x] \p(X(u) = x) \notag \\
			& = \sum_{x\in \cX} \nu_A(\theta)_x\E[e^{\theta A(u, v)}~|~X(u) = x] \pi_x \notag\\
			& = \sum_{x\in \cX} \pi_x\nu_A(\theta)_x \sum_{y\in \cX}\frac{\pi_y}{\pi_x} (\psi(\theta)^{v-u})_{y, x}  \notag \\
			& = \sum_{y\in \cX} \pi_y  \big(\sum_{x\in \cX} \nu_A(\theta)_x\psi(\theta)_{y, x}^{v-u}\big)  \notag \\
			& = \sum_{y\in \cX} \pi_y \nu_A(\theta)_y e^{\theta \rho_A(\theta)(v-u)} \label{eq:vp}\\
			& = e^{\theta \rho_A(\theta)(v-u)}. \label{eq:norm}
		\end{align}
		In the first lines, we use the Markovian assumptions and the time-reversed process construction together with Equation~\eqref{eq:ksteps}. In lines~\eqref{eq:vp} and~\eqref{eq:norm}, we use the fact that $\nu_A(\theta)$ is an eigenvector of $\psi(\theta)$ associated to the eigenvalue $e^{\theta\rho_A(\theta)}$ such that $\langle \pi, \nu_A(\theta) \rangle = 1$. 
}

	\paragraph{Proof of Theorem~\ref{th:martingale}.}
	First, 
	$Y$ and $\sup_{\tau\geq 0} W_{\tau}$ are independent, so from Lemma~\ref{lem:indep_chernoff}, we then only need to show that for all $b\geq 0$, $$\p(\sup_{\tau\geq 0} W_\tau \geq b) \leq \xi_{(A_i), S}(\theta) e^{-\theta b} e^{\theta  (\sum_{i=1}^m \rho_{A_i}(\theta)(v_i-u_i) - \rho_{S}(\theta)(v_0 - u_0))}.$$
	
	Second, as the processes $(A_i)_{i=1}^m$ and $S$ are mutually independent, the martingales $(M_{A_i}(\theta, u_i-\tau, v_i)_{\tau \in\N} )_{i=1}^m$ and $(M_S(\theta, u_0-\tau, v_0))_{\tau \in\N}$ are also mutually independent and the product $(M(\theta, \tau))_{\tau \in \N} = ((\prod_{i=1}^m M_{A_i}(u_i -\tau, v_i)) M_S(u_0-\tau, v_0))_{\tau \in \N}$ is also a martingale with respect to the filtration $(\cF(\tau))_{\tau \in\N} = (\cF_{A_1(u_1-\tau, v_1)}\times \cdots \times \cF_{A_m(u_m-\tau, v_m)}\times \cF_{S(u_0-\tau, v_0)})_{\tau \in\N}$.  
	
	The rest of the proof mainly follows the lines of the proof of  Theorem~3 of~\cite{Duf94}. First, one can decompose the event $\{\sup_{\tau \geq 0} W_\tau \geq b\}$ into elementary events: 
	\begin{align*}
		\{\sup_{\tau \geq 0} W_\tau \geq b\}
		& =  \bigcup_{\tau \geq 0}\{ W_\tau \geq b,  W_{\tau-1} < b\}\\
		& \subseteq  \bigcup_{\tau \geq 0}\{ W_\tau \geq b, \sum_{i=1}^m a_i(u_i - \tau) -  s(u_0 - \tau) > 0 \}.
	\end{align*}
	
	Let us introduce the notations $\cX = \cX_{A_1} \times \cdots \times \cX_{A_m} \times \cX_S$, 
	$X(\tau) = ((X_{A_i}(u_i-\tau))_{i=1}^m, X_{S}(u_0 - \tau))$ and $\nu(\theta)_{X(\tau)} =  \big(\prod_{i=1}^m \nu_{A_i}(\theta)_{X_{A_i}(u_i-\tau)}\big) \cdot  \nu_S(-\theta)_{X_S(u_0-\tau)}$.  
	
	For $\theta\in\R_+$ such that $\sum_{i=1}^m \rho_{A_i}(\theta)  \leq \rho_{S}(\theta)$ (this property is used to obtain the last line), we have
{\small 	\begin{align*}
		\{W_\tau \geq b\} & \subseteq \{e^{\theta [\sum_{i=1}^m A_1(u_i - \tau, v_i) - S(u_0 - \tau, v_0)]} \geq e^{\theta b}\}\\
		&\subseteq \{ M(\theta, \tau) \geq e^{\theta b} \nu(\theta)_{X(\tau)} e^{-\theta(\sum_{i=1}^m \rho_{A_i}(\theta)(v_i-u_i + \tau) - \rho_{S}(\theta)(v_0 - u_0 + \tau))} \}\\
		& \subseteq \{ M(\theta, \tau)\geq e^{\theta b} \nu(\theta)_{X(\tau)}  e^{-\theta(\sum_{i=1}^m \rho_{A_i}(\theta)(v_i-u_i) - \rho_{S}(\theta)(v_0-u_0))} \}.
	\end{align*}}
	Therefore, denoting $c(\tau, X(\tau)) = \sum_{i=1}^m a_i(u_i - \tau) -  s(u_0 - \tau)$, and $\mu(\theta) = \rho_{S}(\theta)(v_0-u_0) - \sum_{i=1}^m \rho_{A_i}(\theta)(v_i-u_i)$,
{\small 	\begin{align*}
		\{W_{\tau} \geq b, c(\tau, X(\tau)) > 0 \} & \subseteq \{ M(\theta, \tau) \geq e^{\theta b} \nu(\theta)_{X(\tau)}  e^{\theta \mu(\theta)},  c(\tau, X(\tau)) > 0\} \\ 
		& \subseteq \Big\{\!M(\theta, \tau) \geq e^{\theta (b+ \mu(\theta))} (\inf_{x\in\cX}\{\nu(\theta)_x|\p(c(\tau, x)>0) > 0\})\!  \Big\}.
	\end{align*}}
	In the last line, we bound $\nu(\theta)_{X(\tau)}$ by the minimum over all possible states $x$, on the condition that there can be more arrivals than services in that state, thanks to the event that $c(\tau, X(\tau)) > 0$. Let us denote 
	$\xi_{(A_i), S}(\theta) = (\inf\{\nu(\theta)_x~|~\p(c(\tau, x)>0) > 0\})^{-1}$.
	
	Using the maximal inequality for positive supermartingales, 	
	\begin{align*}\p(\sup_{\tau \geq 0} W_\tau~|~\cF(0)) & \leq \p\big(\sup_{\tau \geq 0} M(\theta, \tau) \geq \frac{e^{\theta (b+ \mu(\theta)}}{ \xi_{(A_i), S}(\theta) } ~|~ \cF(0)\big)\\
		&\leq  \E[M(\theta, 0)~|~\cF(0)]\xi_{(A_i), S}(\theta)   e^{-\theta (b+ \mu(\theta))},
	\end{align*}
	so $$\p(\sup_{\tau \geq 0} W_\tau) \leq \E_{\pi} [M(\theta, 0)] \xi_{(A_i), S}(\theta)   e^{-\theta (b+ \mu(\theta))},$$
	where $\E_\pi$ is the expectation conditionally to $X(0)$ being distributed as $\pi$, the stationary distribution.
	It remains to compute $\E_{\pi}[M(\theta, 0)]$.
	By independence of the arrival and service processes,
	\begin{align*}
		\E_{\pi}[M(\theta, 0)] & = \prod_{i=1}^m \E_{\pi_{A_i}}[M_{A_i}(u_i, v_i)] \cdot \E_{\pi_S}[M_S(u_0, v_0)].  
	\end{align*}
	For each arrival process (and similar computations can be done for the service process), we have from Lemma~\ref{lem:exp}, and since the processes are stationary
	\begin{align*}
		\E_{\pi_{A_i}}[M_{A_i}(\theta, u_i, v_i)] & =  \E_{\pi_{A_i}}[\nu_{A_i}(\theta)_x e^{\theta A_i(u_i, v_i) - \rho_{A_i}(v_i-u_i)}]\\ 
		& = e^{-\theta\rho_{A_i}(v_i-u_i)} \E[\nu_{A_i}(\theta)_x e^{\theta A_i(u_i, v_i)}] = 1.
	\end{align*}
	Finally, 
	we get $$\p(\sup_{\tau \geq 0} W_\tau) \leq \xi_{(A_i)_{i}, S}(\theta)
	e^{-\theta (b+ \mu(\theta))},$$

	\subsection{Proof of Theorem~\ref{th:main}}
	\label{app:main}
	
	From Theorem~\ref{th:e2e-sc}, an end-to-end dynamic server for flow 1 is
	$$S_{e2e}(t_1, t_{n+1}) = \Big[\inf_{\forall j,~t_j \leq t_{j+1}} \sum_{j=1}^n S_j(t_j, t_{j+1}) - \sum_{i=2}^n A_i(t_{f_i}, t_{\ell_i+1})\Big]_+.$$
	
	From Theorem~\ref{th:bounds}, one can express the violation of the backlog bound $b$ and the delay bound $T$ as
	
	\begin{align*}
		q(t_{n+1}) \geq b & \Rightarrow \sup_{t_1 \leq t_{n+1}} A_1(t_1, t_{n+1}) - S_{2e2}(t_1, t_{n+1}) \geq b \\ &  \Rightarrow \sup_{\substack{1\leq j\leq n \\ t_j \leq t_{j+1}}} ~\sum_{i=1}^m A_i(t_{f_i}, t_{\ell_i+1}) - \sum_{j = 1}^n S_j(t_j, t_{j +1})\geq b,
	\end{align*}
	and 
	\begin{align}
		d(t_{n+1} - T + 1) \geq T
		&\Rightarrow \exists t_1\leq t_{t+1} - T,~A_1(t_1, t_{n+1} - T + 1) > S_{e2e}(t_1, t_{n+1}) \notag\\ 
		&\Rightarrow \exists t_1 \leq t_{n+1}-T, \exists t_1\leq t_2 \leq \cdots \leq t_{n+1},~ \label{eq:delay}\\ &\hspace{0cm}A_1(t_1, t_{n+1} - T + 1)  > \sum_{j=1}^n S_j(t_{j}, t_{j+1}) - \sum_{i=2}^m A_i(t_{f_i}, t_{\ell_ i + 1}).\notag
	\end{align} 
	Let us assume that $(H_6)$ and $(H_7)$ hold for server $h$. 
	For fixed values of  $t_1 \leq  \cdots \leq t_{n+1}$, let us define the new notations $k_1, \ldots, k_{h-1}, \tau$ as:  
	\begin{itemize}
		\item $k_j = t_{j+1} - t_j$ for all $j<h$;
		\item $\tau = t_{h+1} - t_h$.
	\end{itemize}
	
\begin{figure}[htbp]
		\centering
		\begin{tikzpicture}
			\draw[->] (0.5, 0) -- (12, 0);
			\draw (1, 0.1) -- (1, -0.1) node[pos=1, below] {$t_1$};
			\draw (2.5, 0.1) -- (2.5, -0.1) node[pos=1, below] {$t_2$};
			\draw (4, 0.1) -- (4, -0.1) node[pos=1, below] {$t_3$};
			\draw (5.5, 0.1) -- (5.5, -0.1) node[pos=1, below] {$t_{h-1}$};
			\draw (7, 0.1) -- (7, -0.1) node[pos=1, below] {$t_h$};
			\draw (8, 0.1) -- (8, -0.1) node[pos=1, below] {$t_{h+1}$};
			\draw (9, 0.1) -- (9, -0.1) node[pos=1, below] {$t_{h+2}$};
			\draw (11.5, 0.1) -- (11.5, -0.1) node[pos=1, below] {$t_{n+1}$};
			\draw[<->] (1, 0.2) -- (2.5, 0.2) node[pos=0.5, above] {$k_1$};
			\draw[<->] (2.5, 0.2) -- (4, 0.2) node[pos=0.5, above] {$k_2$};
			\draw[<->, dotted] (4, 0.2) -- (5.5, 0.2);
			\draw[<->] (5.5, 0.2) -- (7, 0.2) node[pos=0.5, above] {$k_{h-1}$};
			\draw[<->] (7, 0.2) -- (8, 0.2) node[pos=0.5, above] {$\tau$};
		\end{tikzpicture}
		\caption{Variable change.}
		\label{fig:chgtvar}
	\end{figure}
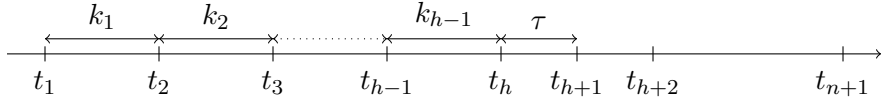

	Using this transformation between variables, we have the equivalence between the two sets 
	$\{(t_1, \ldots, t_{n+1})~|~t_1 \leq \cdots \leq t_{n+1}\}$ and $\{(k_1, \ldots, k_{h-1}, \tau, t_{h+1}, \ldots, t_{n+1})~|~k_1, \ldots, k_{h-1}, \tau \in\N, t_{h+1} \leq \cdots \leq t_{n+1}\}$ (see Figure~\ref{fig:chgtvar}). 
	
	We use the notation $k_j^{j'} = k_j + \cdots + k_{j'}$, with the convention $k_{j}^{j'} = 0$ if $j'< j$.
	For example, for a flow $i$ crossing server $h$, we have
	$$t_{f_i} = t_{h+1} - \sum_{j=f_i}^{h} (t_{j+1} - t_j) = t_{h+1} - \sum_{j=f_i}^{h-1} k_j - \tau = t_{h+1} - k_{f_i}^{h-1} - \tau.$$

	Let us also define the sets $\cL = \{(k_1, \ldots, k_{h-1}, t_{h+1}, t_{n+1})\in \N^{n}~|~t_{h+1} \leq \cdots \leq t_{n+1}\}$ and $\Theta = [0, \theta^*_h] \cap [0, \inf_{j\neq h} \theta^*_j)$.
	
	Each computation in the next paragraphs has the main three steps. 
	\begin{enumerate}
		\item For each  element $K$ of $\cL$, we apply Theorem~\ref{th:martingale}.
		\item We sum all the terms obtained (union bound).
		\item We rewrite and simplify the formula.
	\end{enumerate}
	
	For the sake of simplifying the formulas for the end of the proof, we introduce the following notations to group together the quantities related to servers: 
	\begin{itemize}
		\item  $A_{[h]} = (A_i)_{i\in\fl(h)}$,  $\sigma_{A[-h]} = \sum_{i\notin\fl(h)} \sigma_{A_i}$, $\sigma_{S_{>h}} = \sum_{j>h} \sigma_{S_j}$;
		\item  for each server $j$,  $\rho_j = \rho_{S_j}-\sum_{i\in\fl(j)} \sigma_{A_i} $, $\rho'_j = \rho_{S_j}- \sum_{i\in\fl(j)\setminus \{1\}} \rho_{A_i}$.
	\end{itemize}
For the sake of concision, we will also omit the dependence in $\theta$ of the parameters $\rho$, $\sigma$ and~$\xi$. 
	
	\subsubsection{Backlog}
	\paragraph{Step 1: Application of Theorem~\ref{th:martingale}.}
	For all $\cT = (k_1,\ldots, k_{h-1}, t_{h+1}, \ldots t_{n+1}) \in\cL$, let us define 
	\begin{align*}
		W_{\tau}^{\cT} & = \sum_{i\in \fl(h)} A_i(t_{f_i}, t_{\ell_i + 1}) - S_h(t_h, t_{h+1})\\
		& =  \sum_{i\in \fl(h)} A_i(t_{h+1} - k_{f_i}^{h-1} - \tau, t_{\ell_i + 1}) - S_h(t_{h+1} - \tau, t_{h+1}),
	\end{align*} 
	and the random variable 
	\begin{align*}
		Y^{\cT} & = b + \sum_{j\neq h} S_j(t_j, t_{j+1}) - \sum_{i\notin \fl(h)} A_i(t_{f_i}, t_{\ell_i+1})\\
		& =  b + \sum_{j <  h} k_j C_j + \sum_{j >  h} S_j(t_j, t_{j+1})- \sum_{i\notin \fl(h)} A_i(t_{f_i}, t_{\ell_i+1}). 
	\end{align*}
	As $(H_7)$ holds, for all flow $i\notin \fl(h)$, $f_i > h$. Moreover processes $(A_i)_i$ and $(S_j)_j$ are mutually independent. Therefore, $Y^{\cT}$ is independent of $W_{\tau}^{\cT}$ and Theorem~\ref{th:martingale} can be applied: for all $\theta$ such that $\rho_h(\theta) = \rho_{S_h}(\theta) - \sum_{i\in \fl(h)} \rho_{A_i}(\theta) \geq 0$, there exists $\xi_{A[h], S_h}(\theta)$ such that 
	\begin{align*}
		\p(\sup_{\tau\geq 0} W_\tau^{\cT} \geq Y^{\cT}) & \leq \xi_{A[h], S_h} \E[e^{-\theta Y^{\cT}}] e^{\theta \sum_{i\in \fl(h)} \rho_{A_i}(t_{\ell_i + 1} - t_{h+1} + k_{f_i}^{h-1})}\\
	&  \hspace{-1cm}
		\leq  \xi_{A[h], S_h} e^{-\theta b} e^{\theta \sum_{i\in \fl(h)} \rho_{A_i}(t_{\ell_i + 1} - t_{h+1} + k_{f_i}^{h-1})} \prod_{j=1}^{h-1} e^{-\theta C_jk_j} \\ &\hspace{-0.5cm} \cdot \prod_{j=h+1}^n e^{\theta(\sigma_{S_j} - \rho_{S_j} (t_{j+1} - t_j))} \prod_{i\notin \fl(h)} e^{\theta(\sigma_{A_i}  + \rho_{A_i} (t_{\ell_i+1} - t_{f_i}))} \\
		& \hspace{-1cm} = \xi_{A[h], S_h} e^{\theta (\sigma_{S_{>h}} +  \sigma_{A_{[-h]}})}  e^{-\theta b}  \prod_{j=1}^{h-1} e^{-\theta C_jk_j} \\ & \hspace{-0.5cm} \cdot \prod_{i\in\fl(h)} e^{\theta\rho_{A_i}k_{f_i}^{h-1}}  \prod_{j=h+1}^n e^{-\theta(\rho_{S_j}(t_{j+1} - t_j))} \prod_{i=1}^m e^{\theta \rho_{A_i} (t_{\ell_i+1} - t_{f_i\lor (h+1)})}\\
		& \hspace{-1cm} = \xi_{A[h], S_h} e^{\theta (\sigma_{S_{>h}} +  \sigma_{A_{[-h]}})}  e^{-\theta b}  \prod_{j=1}^{h-1} e^{-\theta \rho_j k_j}\prod_{j=h+1}^n e^{-\theta \rho_j(t_{j+1} - t_j))}.
	\end{align*}
To obtain the last equality, we combine all the contributions that include $t_{j+1} - t_j$ or $k_j$, for each server $j$. That consists of factors related to the service server $j$ itself and the arrivals of the flows crossing that server. 
	
	\paragraph{Step 2: Union bound.}
	Using the right equality of Equation~\eqref{eq:chauchy_n}, 
	we obtain the violation probability, for all $\theta \in \Theta$, 
	
	\begin{align*}
		\p(q(t) \geq b) & \leq \p(\exists \cT \in\cL,~ 
		\sup_{\tau\geq 0} W_\tau^{\cT} \geq Y^{\cT}) \\ \ub  & 
		\leq\sum_{\cT\in\cL}
		\p(\sup_{\tau\geq 0} W_\tau^{\cT} \geq Y^{\cT}) \\  
	{\scriptstyle\text{\textbf{Eq.~\eqref{eq:chauchy_n}}}} 	& \leq   \xi_{A[h], S_h}  e^{\theta (\sigma_{S_{>h}} +  \sigma_{A_{[-h]}})} e^{-\theta b}    \cdot \prod_{j\neq h} \frac{1}{1-e^{-\theta \rho_j}}.
	\end{align*}
	
	\paragraph{Step 3: Rewriting terms.}
	The end-to-end service bgf of flow 1 (cf. Theorem~\ref{th:e2e-bg}) for the tandem network where server $h$ has been removed is: 
	$$F_{S_{e2e^{(-h)}}}(\theta, z) = e^{\theta(\sum_{i\notin H\cup\{1\}} \sigma_{A_i} + \sum_{j\neq h} \sigma_{S_j})} \prod_{j\neq h}\frac{1}{1 - e^{-\theta \rho'_j}z}, $$
	where $H$ is the set of flows crossing server $h$ only. 
	As a consequence, 
	$$F_{S_{e2e}^{(-h)}}(\theta, e^{\theta \rho_{A_1}}) = e^{\theta(\sum_{i\notin H\cup\{1\}} \sigma_{A_i} + \sum_{j\neq h} \sigma_{S_j})} \prod_{j\neq h}\frac{1}{1 - e^{-\theta \rho_j}}.$$
	Finally, including $F_{S_{e2e}^{(-h)}}(\theta, e^{\theta \rho_{A_1}})$ in the formulation leads to the desired result:
	$$\p(q(t) \geq b) \leq \frac{\xi_{A_{[h]}, S_h}e^{-\theta \sigma_{A_1}}}{e^{\theta( \sum_{i\in \fl(h) \setminus H} \sigma_{A_i})}} F_{S_{e2e}^{(-h)}}(\theta, e^{\theta \rho_{A_1}}) e^{-\theta b}.$$ 
	
	\subsubsection{Delay}
	From Equation~\eqref{eq:delay}, the inequality $t_1 \leq t_{n+1} - T$ is equivalent to $\tau\geq t_{h+1}-t_{n+1} -k_1^{h-1}+ T$ and one can write 
	
	\begin{align*}
		d(t_{n+1} - T + 1) \geq &\Rightarrow \exists  k_1,\ldots, k_{h-1}\geq 0,~t_{h+1} \leq \cdots \leq t_{n+1},\\ & \sup_{\tau\geq [t_{h+1} - t_{n+1} +T -k_1^{h-1}]_+} \Big( A_1(t_{h+1} - k_1^{h-1} - \tau, t_{n+1} - T +1)  +  \\ &  \sum_{i\in \fl(h), i\neq 1} A_i(t_{h+1} - k_{f_i}^{h-1} -\tau, t_{\ell_ i + 1})  - S_{h}(t_{h+1} - \tau, t_{h+1}) \Big) \geq \\ &  \sum_{j<h-1} C_jk_j + \sum_{j>h+1} S_j(t_{j}, t_{j+1}) - \sum_{i\notin \fl(h)} A_i(t_{f_i}, t_{\ell_ i + 1}).
	\end{align*} 
	
	\paragraph{Step 1: Application of Theorem~\ref{th:martingale}.}
	With $t' =  [t_{h+1} - t_{n+1} +T -k_1^{h-1}]_+$ and $\tau$ replaced by $\tau' + t'$, let us define for $\cT = (k_1,\ldots, k_{h-1}, t_{h+1},$ $ \ldots, t_{n+1}) \in\cL$ and $\tau' \geq 0$, 
	\begin{multline*}
		W_{\tau'}^{\cT} = A_1(t_{h+1} - k_1^{h-1} - t' - \tau', t_{n+1} - T + 1) \\  +  \sum_{i\in \fl(h)\setminus\{1\}} A_i(t_{h+1} - k_{f_i}^{h-1} -t'- \tau', t_{\ell_ i + 1})   - S_{h}(t_{h+1} - t'- \tau', t_{h+1}),
	\end{multline*}
	and 
	$$Y^{\cT} = \sum_{j<h-1} C_jk_j + \sum_{j>h+1} S_j(t_{j}, t_{j+1}) - \sum_{i\notin \fl(h)} A_i(t_{f_i}, t_{\ell_ i + 1}).$$
	The random variable $Y^{\cT}$ and the process $(W_{\tau'}^{\cT})_{\tau' \geq 0}$ are independent, so by applying Theorem~\ref{th:martingale}, for all $\theta$ such that  $\rho_h(\theta) = \rho_{S_h}(\theta) - \sum_{i\in \fl(h)} \rho_{A_i}(\theta) \geq 0$, 
	there exists $\xi_{A_{[h]}, S_h}(\theta)$ such that 
	\begin{multline*}
		\p(\sup_{\tau'\geq 0} W_{\tau'}^{\cT} \geq Y^{\cT}) \leq \xi_{A_{[h]}, S_h} \E[e^{-\theta Y^{\cT}}] \\ e^{\theta (\rho_{A_1}(t_{n+1} -t_{h+1} + k_1^{h-1} + t' - T + 1) + \sum_{i\in\fl(h)\setminus\{1\}} \rho_{A_i}(t_{\ell_{i+1}} - t_{h+1} + k_{f_i}^{h-1} + t') - \rho_{S_h}t')}.
	\end{multline*}	
	If $t'= 0$, this gives
	\begin{multline*}
		\p(\sup_{\tau'\geq 0} W_{\tau'}^{\cT} \geq Y^{\cT}) \leq \xi_{A_{[h]}, S_h}(\theta) \E[e^{-\theta Y^{\cT}}] \\ e^{\theta (\rho_{A_1}(t_{n+1} -t_{h+1} + k_1^{h-1} - T + 1) + \sum_{i\in\fl(h)\setminus\{1\}} \rho_{A_i}(t_{\ell_{i+1}} - t_{h+1} + k_{f_i}^{h-1}))},
	\end{multline*}	
	and if $t'= t_{h+1} - t_{n+1} +T -k_1^{h-1}$, then
	\begin{multline*}
		\p(\sup_{\tau'\geq 0} W_{\tau'}^{\cT} \geq Y^{\cT}) \leq \xi_{A_{[h]}, S_h}(\theta) \E[e^{-\theta Y^{\cT}}] e^{\theta\rho_{A_1}(\theta)}\\ e^{\theta (\sum_{i\in\fl(h)\setminus\{1\}} \rho_{A_i}(t_{\ell_{i+1}} - t_{n+1} +T -k_1^{f_i-1}) - \rho_{S_h}(t_{h+1} - t_{n+1} +T -k_1^{h-1}))}.
	\end{multline*}

	Now, one can express the violation probability of the delay and separate the terms of the union bound into two groups: 
	\begin{align}
		\p(d(t_{n+1} - T + 1) \geq T) & \leq \p(\exists \cT\in\cL,~ \sup_{\tau'\geq 0} W_{\tau'}^{\cT} \geq Y^{\cT}) \notag \\
		& \leq \sum_{\cT\in\cL } \p(\sup_{\tau'\geq 0} W_{\tau'}^{\cT} \geq Y^{\cT})  \notag\\ 
		& \leq  \sum_{\cT\in\cL, t'= 0}  \p(\sup_{\tau'\geq 0} W_{\tau'}^{\cT} \geq Y^{\cT}) \label{eq:delay_ub} \\ & \hspace{2cm}+\sum_{\cT\in\cL, t'> 0} \p(\sup_{\tau'\geq 0} W_{\tau'}^{\cT} \geq Y^{\cT}). \notag
	\end{align}	
	
	\paragraph{Step 2a: Union bound [$t'= 0$].}
	
	Let us first give a bound on the left-hand sum term of Equation~\eqref{eq:delay_ub}. Having $t'= 0$ is equivalent to $t_{n+1} - t_{h+1} +k_1^{h-1} -T \geq 0$.  In line~\eqref{eq:del1}, we  sum over all possible values $u$; in line~\eqref{eq:delay1}, we regroup the terms per servers and use $k_j = t_{j+1} - t_j$;  and in line~\eqref{eq:del2}, we recognize the term of the end-to-end service bgf using Equation~\eqref{eq:chauchy_n}. 
	For all $\theta_1\in \Theta$, 
	\begin{align}
		P_1(\theta_1)& = \sum_{\cT\in \cL, t_{n+1} - t_{h+1} + k_1^{h-1} - T \geq 0} \p(W_\tau^{\cT} \geq Y^{\cT}) \notag \\
		& \leq \sum_{u \geq 0}  \xi_{A_{[h]}, S_h} e^{\theta_1 (\rho_{A_1}(u + 1) +  \sigma_{A_{[-h]}} + \sigma_{S_{>h}})} \label{eq:del1}\\
		& \hspace{-1cm}\cdot \sum_{\cT \in\cL, t_{n+1} - t_{h+1} + k_1^{h-1} - T = u} \hspace{-1.5cm}  e^{\theta_1 (\sum_{i=2}^m \rho_{A_i}(t_{\ell_i + 1} - t_{f_i\lor h + 1} - k_{f_i}^{h-1}) - \sum_{j<h-1}^n C_jk_j - \sum_{j>h} \rho_{S_j}(t_{j+1} - t_j))} \notag\\
		& = \sum_{u \geq 0}  \xi_{A_{[h]}, S_h} e^{\theta_1 (\rho_{A_1}(u + 1) +  \sigma_{A_{[-h]}} + \sigma_{S_{>h}})} \cdot \sum_{\sum_{j\neq h} k_j = u + T} \prod_{j\neq h} e^{\theta_1 \rho'_j k_j}  \label{eq:delay1} \\
		& \leq \sum_{u \geq 0} \xi_{A_{[h]}, S_h} e^{\theta_1 (\rho_{A_1} (u+1) - \sum_{i\in\fl(h) \setminus(H\cup\{1\})} \sigma_{A_i})} 
		[z^{u + T}] F_{S_{e2e}}^{(-h)} (\theta_1, z). \label{eq:del2}
	\end{align}

	\paragraph{Step 3a: Rewriting terms [$t'= 0$].}
	One can now notice that $$\frac{e^{\theta_1 \sum_{i\in\fl(h) \setminus H} \sigma_{A_i}}}{\xi_{A_{[h]}, S_h}} P_1(\theta_1) \leq \sum_{u \geq 0}  e^{\theta_1 (\sigma_{A_1 }+ \rho_{A_1} (u+1))} 	[z^{u + T}] F_{S_{e2e}}^{(-h)} (\theta_1, z),$$
	and we recognize the right-hand side as the $T$-th term of the delay bounding generating function for the arrival process $A_1$ and dynamic $S_{e2e}^{(-h)}$-server according to~\cite[Eq.(11)]{BNS22a}, 
	so for all $\theta_1 \in \Theta$,
	$$P_1(\theta_1) \leq  \frac{\xi_{A_{[h]}, S_h}(\theta_1)}{e^{\theta_1 \sum_{i\in\fl(h) \setminus H} \sigma_{A_i}(\theta_1)}} [z^{T}] F_{d, S_{e2e}^{(-h)}}(\theta_1, z).$$

	\paragraph{Step 2b: Union bound [$t'> 0$].}
	Let us now focus on the right-hand sum term of Equation~\eqref{eq:delay_ub}: if $t'>0$, then one must have $t_{n+1} - t_{h+1} + k_1^{h-1} < T$, and the only possible values for $t'$ are  $\{1, \ldots, T\}$. Rewriting the third line with $k_h = t'$ and $k_j = t_{j+1} - t_j$ for $j>h$, and $C_j = \rho_j(\theta)$, we recognize the expression of the product of geometric series of Equation~\eqref{eq:chauchy_n} and one gets for all $\theta_2\in[0, \theta^*_h]$, 
	\begin{align*}
		P_2(\theta_2) & = \sum_{\cT\in\cL, t'>0} \p(\sup_{\tau\geq 0} W_\tau^{\cT} \geq Y^{\cT}) \\
		&\leq \sum_{\substack{t_{n+1} -t_{h+1} + k_1^{h-1} + t' = T\\ (k_1,\ldots, t_{n+1}) \in \cL}}  \hspace{-0.5cm}
		 \xi_{A_{[h]}, S_h} e^{\theta_2 (\sum_{i\in\fl(h)\setminus\{1\} }\rho_{A_i}(t_{\ell_i + 1} - t_{h+1} + k_{f_i}^{h-1} + t')  - \rho_{S_h}t')}\\ 
		& e^{\theta_2(\rho_{A_1} + \sigma_{A_{[-h]}} + \sigma_{S_{>h}})} e^{\theta_2 (\sum_{i\notin \fl(h)} \rho_{A_i}(t_{\ell_i + 1} - t_{f_i}) - \sum_{j <h} C_jk_j - \sum_{j>h}  \rho_{S_j}(t_{j+1} - t_j))}\\
		& = \xi_{A_{[h]}, S_h} e^{\theta_2(\rho_{A_1} + \sigma_{A_{-[h]}} + \sigma_{S_{>h}})}  \sum_{\substack{k_j \geq 0, k_h>0\\ \sum_j k_j = T}}   e^{-\theta_2 (\sum_{j = 1}^n  \rho'_j k_j)}\\
		&\leq \xi_{A_{[h]}, S_h} e^{\theta_2(\rho_{A_1} +\sigma_{A_{[-h]}} + \sigma_{S_{>h}})} [z^{T-1}]  \prod_{j=1}^n\frac{ e^{-\theta_2 \rho'_h} }{ 1- e^{\theta_2 \rho'_j}z}.
	\end{align*}
	
	\paragraph{Step 3b: Rewriting terms [$t'> 0$].}
	
	The bounding generating function for the end-to-end service curve for flow 1 is 
	$$F_{S_{e2e}}(\theta_2, z) = e^{\theta_2(\sum_{i=2}^n \sigma_{A_i} + \sum_{j\geq h} \sigma_{S_j})} \prod_{j=1}^n \frac{1}{1- e^{\theta_2 \rho'_j}z}, 
	$$
	so 
	$$P_2(\theta_2) \leq \frac{\xi_{A_{[h]}, S} e^{-\theta_2(\rho_{S_h} - \sum_{i\in\fl(h)} \rho_{A_i})}}{e^{\theta_2(\sigma_{S_h} + \sum_{i\in\fl(h)\setminus\{1\}}\sigma_{A_i})}}[z^{T-1}] F_{S_{e2e}}(\theta_2, z).$$
	Noticing that $[z^{T-1}] f(z) = [z^T] zf(z)$ and summing the two obtained bounds concludes the proof.
	
	\section{Conclusion}
	\label{sec:conclusion}
	In this paper, we have presented a method to compute probabilistic  end-to-end performance bounds in networks that combines two types of analysis from Stochastic Network Calculus: we locally use the martingale analysis at one server and use the more classical MGF-based SNC results. In particular, this can avoid the use of the union bound at the bottleneck of the network and simulations show that, for small networks, the gap between simulation and best results from the state of the art is drastically reduced. 
	
	Of course, the improvement may vanish when considering larger networks, and further investigations are needed to achieve tight bounds. The challenge is to bound the maximum of random variables that cannot be, or have not yet been, expressed as a martingale. 
	
	Another research direction is to investigate other service policies. Indeed,  the  MGF-based SNC has until now mainly focused on blind multiplexing, which encompasses all the possible service policies, hence the more pessimistic for the flow of interest.

\bibliographystyle{plain}

\begin{thebibliography}{10}
	
	\bibitem{AK11}
	Kishore Angrishi and Ulrich Killat.
	\newblock An approach using n-demisupermartingales for the stochastic analysis
	of networks.
	\newblock {\em CoRR}, abs/1111.3063, 2011.
	
	\bibitem{Bec16}
	Michael Beck.
	\newblock {\em Advances in Theory and Applicability of Stochastic Network
		Calculus}.
	\newblock PhD Thesis of the University of Kaiserslautern, 2016.
	
	\bibitem{BBL18}
	Anne Bouillard, Marc Boyer, and Euriell Le~Corronc.
	\newblock {\em {Deterministic Network Calculus: From Theory to Practical
			Implementation}}.
	\newblock ISTE, 2018.
	
	\bibitem{BGLT2008}
	Anne Bouillard, Bruno Gaujal, Sébastien Lagrange, and {\'E}ric. Thierry.
	\newblock {Optimal routing for end-to-end guarantees using Network Calculus}.
	\newblock {\em Performance Evaluation}, 65(11-12):883--906, 2008.
	
	\bibitem{BNS21}
	Anne Bouillard, Paul Nikolaus, and Jens~B. Schmitt.
	\newblock Fully unleashing the power of paying multiplexing only once in
	stochastic network calculus.
	\newblock {\em CoRR}, abs/2104.14215, 2021.
	
	\bibitem{BNS22a}
	Anne Bouillard, Paul Nikolaus, and Jens~B. Schmitt.
	\newblock Unleashing the power of paying multiplexing only once in stochastic
	network calculus.
	\newblock {\em Proc. {ACM} Meas. Anal. Comput. Syst.}, 6(2):31:1--31:27, 2022.
	
	\bibitem{BNS22b}
	Anne Bouillard, Paul Nikolaus, and Jens~B. Schmitt.
	\newblock Unleashing the power of paying multiplexing only once in stochastic
	network calculus.
	\newblock In D.~Manjunath, Jayakrishnan Nair, Niklas Carlsson, Edith Cohen, and
	Philippe Robert, editors, {\em {SIGMETRICS/PERFORMANCE} '22: {ACM}
		{SIGMETRICS/IFIP} {PERFORMANCE} Joint International Conference on Measurement
		and Modeling of Computer Systems, Mumbai, India, June 6 - 10, 2022}, pages
	117--118. {ACM}, 2022.
	
	\bibitem{BN15}
	Anne Bouillard and Thomas Nowak.
	\newblock {Fast symbolic computation of the worst-case delay in tandem networks
		and applications}.
	\newblock {\em Perform. Eval.}, 91:270--285, 2015.
	
	\bibitem{Chang1994}
	Cheng-Shang Chang.
	\newblock Stability, queue length, and delay of deterministic and stochastic
	queueing networks.
	\newblock {\em IEEE Transactions on Automatic Control}, 39(5):913--931, 1994.
	
	\bibitem{Chang2000}
	Cheng-Shang Chang.
	\newblock {\em {Performance Guarantees in Communication Networks}}.
	\newblock TNCS, Springer-Verlag, 2000.
	
	\bibitem{CBL06}
	Florin Ciucu, Almut Burchard, and J{\"o}rg Liebeherr.
	\newblock Scaling properties of statistical end-to-end bounds in the network
	calculus.
	\newblock {\em IEEE/ACM Transactions on Networking (ToN)}, 14(6):2300--2312,
	2006.
	
	\bibitem{CP19}
	Florin Ciucu and Felix Poloczek.
	\newblock Two extensions of kingman's {GI/G/1} bound.
	\newblock In Erich~M. Nahum, Thomas Bonald, and Nick Duffield, editors, {\em
		Abstracts of the 2019 SIGMETRICS/Performance Joint International Conference
		on Measurement and Modeling of Computer Systems}, pages 63--64. {ACM}, 2019.
	
	\bibitem{CPCC21}
	Florin Ciucu, Felix Poloczek, Lydia~Y. Chen, and Martin Chan.
	\newblock Practical analysis of replication-based systems.
	\newblock In {\em 40th {IEEE} Conference on Computer Communications, {INFOCOM}
		2021}, pages 1--10, 2021.
	
	\bibitem{Cruz1995}
	Rene~L. Cruz.
	\newblock {Quality of Service Guarantees in Virtual Circuit Switched Networks}.
	\newblock {\em IEEE Journal on selected areas in communication}, 13:1048--1056,
	1995.
	
	\bibitem{Duf94}
	Nick~G. Duffield.
	\newblock Exponential bounds for queues with markovian arrivals.
	\newblock {\em Queueing Syst. Theory Appl.}, 17(3-4):413--430, 1994.
	
	\bibitem{Fid06}
	M.~{Fidler}.
	\newblock {An End-to-End Probabilistic Network Calculus with Moment Generating
		Functions}.
	\newblock In {\em 14th IEEE International Workshop on Quality of Service},
	pages 261--270, 2006.
	
	\bibitem{FR15}
	Markus Fidler and Amr Rizk.
	\newblock A guide to the stochastic network calculus.
	\newblock {\em IEEE Communications Surveys Tutorials}, 17(1):92--105, 2015.
	
	\bibitem{Kin64}
	John F.~C. Kingman.
	\newblock A martingale inequality in the theory of queues.
	\newblock {\em Mathematical Proceedings of the Cambridge Philosophical
		Society}, 60(2):359–361, 1964.
	
	\bibitem{LT2001}
	Jean-Yves Le~Boudec and Patrick Thiran.
	\newblock {\em {Network Calculus: A Theory of Deterministic Queuing Systems for
			the Internet}}, volume LNCS 2050.
	\newblock Springer-Verlag, 2001.
	\newblock revised version 4, May 10, 2004.
	
	\bibitem{LJ08}
	Yong Liu and Yuming Jiang.
	\newblock {\em {Stochastic Network Calculus}}.
	\newblock Springer, 2008.
	
	\bibitem{NS17-1}
	Paul Nikolaus and Jens Schmitt.
	\newblock {On Per-Flow Delay Bounds in Tandem Queues under (In)Dependent
		Arrivals}.
	\newblock In {\em Proc. 16th IFIP Networking Conference (NETWORKING'17)}, pages
	1--9, Stockholm, Sweden, 2017.
	
	\bibitem{NS20}
	Paul Nikolaus and Jens Schmitt.
	\newblock Improving delay bounds in the stochastic network calculus by using
	less stochastic inequalities.
	\newblock In {\em Proc. 13th EAI International Conference on Performance
		Evaluation Methodologies and Tools (VALUETOOLS 2020)}, VALUETOOLS '20, page
	96–103, New York, NY, USA, 2020. Association for Computing Machinery.
	
	\bibitem{PC14}
	Felix Poloczek and Florin Ciucu.
	\newblock Scheduling analysis with martingales.
	\newblock {\em Performance Evaluation}, 79:56 -- 72, 2014.
	\newblock Special Issue: Performance 2014.
	
	\bibitem{PC15}
	Felix Poloczek and Florin Ciucu.
	\newblock Service-martingales: Theory and applications to the delay analysis of
	random access protocols.
	\newblock In {\em 2015 {IEEE} Conference on Computer Communications, {INFOCOM}
		2015}, pages 945--953, 2015.
	
	\bibitem{Rao07}
	B.~L. S.~Prakasa Rao.
	\newblock On some maximal inequalities for demisubmartingales and n demisuper
	martingales.
	\newblock {\em Journal of Inequalities in Pure \& Applied Mathematics}, 8,
	2007.
	
	\bibitem{SZM08}
	Jens~B. Schmitt, Frank~A. Zdarsky, and Ivan Martinovic.
	\newblock {Improving Performance Bounds in Feed-Forward Networks by Paying
		Multiplexing Only Once}.
	\newblock In Falko Bause and Peter Buchholz, editors, {\em Proceedings 14th
		{GI/ITG} Conference on Measurement, Modelling and Evaluation of Computer and
		Communication Systems {(MMB} 2008), March 31 - April 2, 2008, Dortmund,
		Germany}, pages 13--28. {VDE} Verlag, 2008.
	
	\bibitem{SF96}
	Robert Sedgewick and Philippe Flajolet.
	\newblock {\em An Introduction to the Analysis of Algorithms}.
	\newblock Addison-Wesley Longman Publishing Co., Inc., USA, 1996.
	
	\bibitem{YS93}
	Opher Yaron and Moshe Sidi.
	\newblock Performance and stability of communication networks via robust
	exponential bounds.
	\newblock {\em {IEEE/ACM} Trans. Netw.}, 1(3):372--385, 1993.
	
	\bibitem{YCL22}
	Baozhu Yu, Xuefen Chi, and Xuan Liu.
	\newblock Martingale-based bandwidth abstraction and slice instantiation under
	the end-to-end latency-bounded reliability constraint.
	\newblock {\em IEEE Communications Letters}, 26(1):217--221, 2022.
	
\end{thebibliography}

\end{document}